\documentclass[twocolumn]{aastex63}

\shorttitle{Perkins INfrared Exosatellite Survey}
\shortauthors{Tamburo et al.}

\usepackage{amsmath}

\begin{document}

\title{The Perkins INfrared Exosatellite Survey (PINES) I. Survey Overview, Reduction Pipeline, and Early Results}

\correspondingauthor{Patrick Tamburo}
\email{tamburop@bu.edu}

\author[0000-0003-2171-5083]{Patrick Tamburo}
\affiliation{Department of Astronomy \& The Institute for Astrophysical Research, Boston University, 725 Commonwealth Ave., Boston, MA 02215, USA}

\author[0000-0002-0638-8822]{Philip S. Muirhead}
\affiliation{Department of Astronomy \& The Institute for Astrophysical Research, Boston University, 725 Commonwealth Ave., Boston, MA 02215, USA}

\author[0000-0003-2015-5029]{Allison M. McCarthy}
\affiliation{Department of Astronomy \& The Institute for Astrophysical Research, Boston University, 725 Commonwealth Ave., Boston, MA 02215, USA}

\author[0000-0002-0330-1648]{Murdock Hart}
\affiliation{Department of Astronomy \& The Institute for Astrophysical Research, Boston University, 725 Commonwealth Ave., Boston, MA 02215, USA}
\affiliation{Perkins Telescope Observatory, 2780 Marshall Lake Rd., Flagstaff, AZ 86005, USA}

\author[0000-0002-0447-7231]{David Gracia}
\affiliation{Department of Astronomy \& The Institute for Astrophysical Research, Boston University, 725 Commonwealth Ave., Boston, MA 02215, USA}

\author[0000-0003-0489-1528]{Johanna M. Vos}
\affiliation{American Museum of Natural History, 200 Central Park West, New York, NY 10024, USA}

\author[0000-0001-8170-7072]{Daniella C. Bardalez Gagliuffi }
\affiliation{American Museum of Natural History, 200 Central Park West, New York, NY 10024, USA}

\author[0000-0001-6251-0573]{Jacqueline Faherty}
\affiliation{American Museum of Natural History, 200 Central Park West, New York, NY 10024, USA}

\author[0000-0002-9807-5435]{Christopher Theissen}
\affiliation{Center for Astrophysics and Space Studies, University of California, San Diego, La Jolla, CA 91125, USA}
\affiliation{NASA Sagan Fellow}

\author[0000-0002-0802-9145]{Eric Agol}
\affiliation{Department of Astronomy \& Virtual Planetary Laboratory, University of Washington, Box 351580, U.W., Seattle, WA 98195, USA}

\author[0000-0002-4794-1591]{Julie N. Skinner}
\affiliation{Department of Astronomy \& The Institute for Astrophysical Research, Boston University, 725 Commonwealth Ave., Boston, MA 02215, USA}

\author[0000-0002-3022-6858 ]{Sheila Sagear}
\affiliation{Department of Astronomy, University of Florida, Gainesville, FL 32611, USA}

\begin{abstract}

We describe the Perkins INfrared Exosatellite Survey (PINES), a near-infrared photometric search for short-period transiting planets and moons around a sample of 393 spectroscopically confirmed L- and T-type dwarfs. PINES is performed with Boston University's 1.8 m Perkins Telescope Observatory, located on Anderson Mesa, Arizona. We discuss the observational strategy of the survey, which was designed to optimize the number of expected transit detections, and describe custom automated observing procedures for performing PINES observations. We detail the steps of the \texttt{PINES Analysis Toolkit} (\texttt{PAT}), software that is used to create light curves from PINES images. We assess the impact of second-order extinction due to changing precipitable water vapor on our observations and find that the magnitude of this effect is minimized in Mauna Kea Observatories \textit{J}-band. We demonstrate the validity of \texttt{PAT} through the recovery of a transit of WASP-2 b and known variable brown dwarfs and use it to identify a new variable L/T transition object: the T2 dwarf WISE J045746.08-020719.2. We report on the measured photometric precision of the survey and use it to estimate our transit detection sensitivity. We find that for our median brightness targets, assuming contributions from white noise only, we are sensitive to the detection of 2.5 $R_\oplus$ planets and larger. PINES will test whether the increase in sub-Neptune-sized planet occurrence with decreasing host mass continues into the L and T dwarf regime.
\end{abstract}


\section{Introduction} 
\label{sec:intro}

The L and T spectral types extend the traditional Harvard stellar classification system \citep{Cannon1901} to include objects that are less massive, cooler, and spectroscopically distinct from the latest M dwarfs \citep{Kirkpatrick1999a, Burgasser2006}. With effective temperatures (T$_{eff}$) in the range from $\sim$700 to $\sim$2200 K \citep[e.g.,][]{Vrba2004,Nakajima2004}, L- and T-type dwarfs are faint at optical wavelengths, and only recently began to be detected in appreciable numbers, largely thanks to large-scale near-infrared (NIR) surveys like the the DEep Near Infrared Survey of the Southern Sky \citep[DENIS;][]{Epchtein1997}, the 2-Micron All Sky Survey \citep[2MASS;][]{Skrutskie2006}, the UKIRT Infrared Deep Sky Survey \citep[UKIDSS;][]{Lawrence2007}, and the space-based \textit{Wide-field Infrared Survey Explorer} \citep[\textit{WISE};][]{Wright2010}. Today, thousands of L- and T-type dwarfs are known. Studies of the initial mass function (IMF) have found that though they are less common by number per unit volume than stars, L- and T-type dwarfs are still frequent outcomes of the process of star formation \citep{Kroupa2001,Chabrier2005,Kirkpatrick2012,Kirkpatrick2019}.

The L spectral class spans the boundary between the stellar regime, where sustained hydrogen fusion maintains long main-sequence lifetimes, and the substellar regime of brown dwarfs (BDs) and planets, where the lack of sustained fusion causes objects to cool off over time, radiating away their heat of formation \citep{Kumar1963a,Hayashi1963}. It should be emphasized that the L and T taxonomy constitutes a \textit{spectral} sequence, being defined by the evolution of observable spectral features across the classes; the L spectral class does not represent a demarcation between stars and BDs/planets. Rather, the spectral features that characterize early L dwarfs can equally be achieved in the atmospheres of stars ($\gtrsim 1$ Gyr),  BDs ($\sim0.01 - 1$ Gyr), and planetary-mass objects \citep[$\lesssim 0.01$ Gyr;][]{Burrows1997}. Only by mid-L ($\sim$L4) are objects unambiguously sub-stellar.

Despite advances in the understanding of L and T dwarfs in recent decades, their short-period (P $\lesssim$ 200 days) planet population remains largely unconstrained. The occurrence rates of short-period planets around earlier spectral types have been studied in some detail with optical radial velocity (RV) and transit surveys \citep[e.g.,][]{Johnson2010, Howard2012, Dressing2013, Dressing2015, Mulders2015a, HardegreeUllman2019}, but these surveys generally lack the sensitivity to detect planets around optically faint L and T dwarfs. The NIRSPEC Ultracool Dwarf Radial Velocity Survey \citep{Blake2010} targeted a sample of 59 late-M/early-L dwarfs at NIR wavelengths but was designed to search for giant planets and BD companions. As a result, knowledge about sub-Neptune short-period L/T planet occurrence rates is currently limited to upper limits from \textit{Spitzer} \citep{He2017} and \textit{K2} data \citep{Sagear2020, Sestovic2020}. Individual \textit{long}-period planets have been discovered around BDs through direct imaging \citep[e.g.,][]{Chauvin2004, Close2007, Bejar2008, Luhman2009, Fontanive2020}, gravitational microlensing \citep[e.g.,][]{Han2013,Jung2018a, Jung2018b}, and astrometry \citep[][]{Sahlmann2013}, but those that have been confirmed have large planet-to-host mass ratios and orbital semi-major axes that are generally suggestive of formation through gravitational instability  \citep[e.g.,][]{Chauvin2005}, rather than the core accretion mechanism that is thought to govern the in-situ formation of short-period planets.

There is mixed evidence regarding the potential for planet formation around L and T dwarfs through core accretion. Analyses of \textit{Kepler} data \citep{Borucki2010} have revealed that short-period, sub-Neptune-sized planets occur more frequently with later host spectral types, with M-type stars hosting about three times as many sub-Neptunes as F-type stars \citep{Howard2012, Mulders2015a}. M dwarfs are also frequently observed to host ``compact multiple" systems, containing multiple sub-Neptune-sized planets on orbits with periods less than $\sim$10 days \citep{Muirhead2015,Ballard2016}. TRAPPIST-1, an M8 dwarf, is known to host seven transiting, Earth-sized exoplanets \citep{Gillon2016,Gillon2017}. With a mass of 93 $M_{Jup}$ \citep{VanGrootel2018}, TRAPPIST-1 is just slightly more massive than a field L or T dwarf and was one of only $\sim$50 systems targeted by the TRAPPIST Ultra-Cool Dwarfs Transit Survey, a prototype survey for SPECULOOS \citep{Gillon2013a, Delrez2018}. Such systems may also be common around L or T dwarfs, but a recent pebble accretion model from \citet{Mulders2021} predicts a turn-over in the occurrence rate of super-Earths past the M spectral type, and that the formation of such planets may cease around objects less massive than 0.1 $M_\odot$. 

Disks of gas and dust, the reservoirs from which planets form in the core accretion model, have been observed to be prevalent around BDs, occurring with similar frequencies and dissipation timescales as those around stellar-mass objects \citep{Luhman2005, Luhman2007, Monin2010, Dawson2013, Liu2015}. In some cases, the masses of BD protoplanetary disks have been measured, with dust mass estimates that typically range from a fraction of an Earth mass to a few Earth masses \citep[e.g.,][]{Klein2003,Scholz2006,Harvey2012,Broekhoven-Fiene2014,vanderPlas2016,Rilinger2019}. \citet{Rilinger2021} measured masses for the largest sample of BD protoplanetary disks to date using spectral energy distribution (SED) modeling, and only 8 out of 49 disks in their sample possessed masses greater than 1 $M_{Jup}$. Planet formation simulations suggest that these measured masses are generally insufficient to form Earth-mass planets through core accretion. \citet{Payne2007} performed core accretion modeling for planet formation around BDs and found that disks with masses less than 1 $M_{Jup}$  rarely form planets more massive than 0.3 $M_\oplus$. \citet{Miguel2020} adapted a code built to study the formation of the Galilean satellites to planet formation in BD disks, finding that disks with masses greater than $\sim$10 $M_{Jup}$ were required to form planets more massive than 0.1 $M_\oplus$. However, they note that their modeling fails to explain some of the larger known exoplanets around M dwarfs.

The tension between the competing lines of evidence presented above can only be resolved by searching for short-period planets around L and T dwarfs. With current technology, these planets are most amenable to detection using the transit method, as the diminutive radii of their hosts result in transit depths on the order of 1\% for Earth-sized planets, signals that are accessible with 1- to 2-m class telescopes equipped with NIR photometers \citep{Tamburo2019}. However, transit searches will also have to contend with short-duration transit events \citep[down to 15 minutes, ][]{Delrez2018} and percent-level photometric variability on hours-long timescales that is frequently exhibited by L and T dwarfs \citep{Radigan2014, Wilson2014, Metchev2015, Vos2019, Tannock2021}.

One such transit survey, the Search for habitable Planets EClipsing ULtra-cOOl Stars (SPECULOOS), began in January 2019 and is looking for transiting planets around a sample of $\sim$1200 ultracool dwarfs (UCDs), of which $\sim$13.4\% are L and T dwarfs \citep{Delrez2018, Murray2020,Sebastian2021}. SPECULOOS consists of a worldwide array of robotic telescopes (most of which are 1 m) which perform nightly photometric observations in a custom \textit{I+z'} filter.

In this work, we describe a new effort to search for transiting companions around L and T dwarfs at wavelengths longer than 1.0 $\mu$m: the Perkins INfrared Exosatellite\footnote{Because L and T dwarfs can themselves be planetary mass, the term “exosatellite” is used as a more general substitute for “exoplanet” in the survey title.} Survey (PINES). In Section \ref{sec:survey}, we provide an overview of the survey, describing the observing facility, target sample, and observing strategy. Section \ref{sec:pipeline} details the custom photometric pipeline for analyzing PINES data. In Section \ref{sec:performance}, we quantify the survey's photometric performance and use this to estimate our transit detection sensitivity. We also validate our pipeline's performance using observations of a transiting exoplanet and known variable L/T transition objects and identify a previously unknown variable.

\section{The PINES Survey}
\label{sec:survey}

\subsection{Observing Facility and Instrument}
\label{subsec:facility}
PINES is conducted using Boston University's 1.8 m Perkins Telescope Observatory (PTO), located on Anderson Mesa, Arizona, at an altitude of 2206 m. The telescope is a Cassegrain reflector with a 0.4 m diameter secondary mirror. It is mounted on an English cross-axis equatorial mount originally built in the 1920s \citep{Crump1929}.

In the survey,  we perform photometric observations of L and T dwarfs using Mimir \citep{Clemens2007}, a NIR polarimeter, spectrometer, and imager that is permanently housed at the PTO. We use the instrument in its broadband \textit{J}- and \textit{H}-band wide-field imaging modes which provide 10'x10' field-of-view (FOV) images. The FOV is large enough to provide several similarly bright reference stars in each target field, which are needed to measure and remove effects from changing atmospheric conditions \citep{Tamburo2019}. The \textit{J}- and \textit{H}-band filters are from the Mauna Kea Observatories near-infrared (MKO-NIR) set, which was designed in part to avoid contamination from water vapor lines \citep{Simons2002, Tokunaga2002}.

\begin{figure}[b!]
    \centering
    \includegraphics[width=\columnwidth]{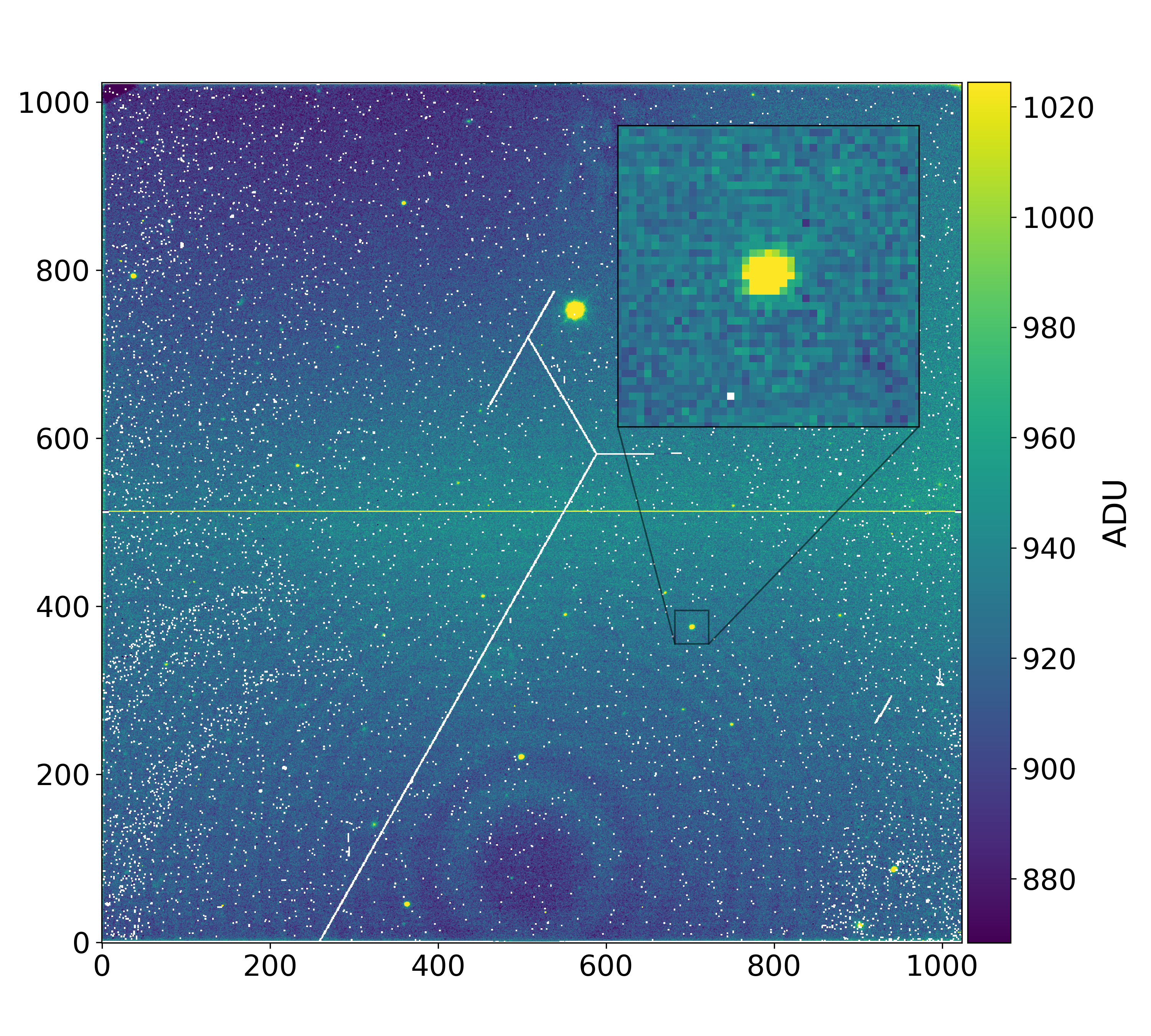}
    \caption{A typical reduced \textit{J}-band Mimir image. Bad pixels in this image are shown in white. The inset shows a zoom-in on the target in this field, which is deliberately placed in a detector region that is relatively devoid of bad pixels.}
    \label{fig:mimir_image}
\end{figure}

The Mimir detector is an ALADDIN III InSb 1024$\times$1024 array detector, which is operated at 33.5 K with the help of a Trillium 1050 two-stage, closed-cycle helium cooling system. A typical reduced Mimir image is shown in Figure \ref{fig:mimir_image}. On average, we find that 2.7\% of the pixels are either significantly hot, dark, or variable, and are thus unsuitable for precise photometry; these pixels are marked in white in Figure \ref{fig:mimir_image}, and we detail how they were identified in Section \ref{subsec:bad_pixel_identification}. Bad pixels are spread more or less homogeneously across the chip, except for in a physical crack that runs through all four detector quadrants. Additionally, Figure \ref{fig:mimir_image} shows an interference pattern in the bottom two quadrants, likely due to sky background interference within one or several refractive optical elements. The resulting interference pattern causes a series of concentric rings of alternating brightness in the lower two quadrants. We modeled the ring pattern to remove it from the reduced images but found that doing so had minimal impact on the resulting light curve quality.

\subsection{Target Sample}
\label{sec:target_sample} 
We assembled the PINES target sample using an online database of ultracool dwarfs maintained by J. Gagn{\'e}\footnote{\url{https://jgagneastro.com/list-of-ultracool-dwarfs/}}, which incorporates targets listed in \citet{Mace2014}, \citet{Dupuy2012}, and the DwarfArchives.org catalog. All objects in the list have either an optical or NIR spectral type of L0 or later. We selected targets from this list that reside between the declination limits of the PTO (-10$^{\circ} < \delta < $ +70$^{\circ}$) and have an apparent 2MASS magnitude of $m_J < 16.5$. Known unresolved binaries were also removed from the list. In total, the PINES sample consists of 393 spectroscopically confirmed L and T dwarf targets, the properties of which are given in Table \ref{tab:targets}.

\begin{table}[t]
\startlongtable
\begin{deluxetable*}{lrrrrrll}
    \label{tab:targets}
    \tablewidth{0pt}
    \tablecaption{Targets in the PINES sample first by group number and then by R.A. Full table available online.}
    \tablehead{
        \colhead{2MASS Name} & \colhead{PINES} & \colhead{R.A.} & \colhead{Dec} & \colhead{2MASS} & \colhead{2MASS} & \colhead{SpT} & \colhead{SpT} \\
        \colhead{} & \colhead{Group ID} & \colhead{(J2000)} & \colhead{(J2000)} & \colhead{m$_J$} & \colhead{m$_H$} & \colhead{Optical} & \colhead{NIR}
    }
    \startdata
        00001354+2554180 & 0 & 00:00:14 & +25:54:17 & 15.063 ± 0.039 & 14.731 ± 0.074 & ... & T4.5 \\
        00154476+3516026 & 0 & 00:15:45 & +35:16:02 & 13.878 ± 0.028 & 12.892 ± 0.035 & L2 & L1 \\
        00193275+4018576 & 0 & 00:19:33 & +40:18:54 & 15.544 ± 0.057 & 14.928 ± 0.068 & ... & L2 \\
        00250365+4759191 & 0 & 00:25:04 & +47:59:19 & 14.840 ± 0.036 & 13.667 ± 0.030 & L4 & ... \\
        00144919-0838207 & 1 & 00:14:50 & -08:38:22 & 14.469 ± 0.026 & 13.950 ± 0.026 & L0 & M9 \\
        00191165+0030176 & 1 & 00:19:12 & +00:30:18 & 14.921 ± 0.035 & 14.180 ± 0.040 & L1 & L0.5 \\
        00242463-0158201 & 1 & 00:24:25 & -01:58:17 & 11.992 ± 0.033 & 11.084 ± 0.020 & M9.5 & L0.5 \\
        00261147-0943406 & 1 & 00:26:11 & -09:43:40 & 15.601 ± 0.067 & 14.999 ± 0.081 & L1 & ... \\
        00320509+0219017 & 1 & 00:32:05 & +02:19:01 & 14.324 ± 0.023 & 13.386 ± 0.023 & L1.5 & M9 \\
        00345684-0706013 & 1 & 00:34:57 & -07:06:00 & 15.531 ± 0.059 & 14.566 ± 0.041 & L3 & L4.5 \\
        00011217+1535355 & 2 & 00:01:12 & +15:35:36 & 15.522 ± 0.061 & 14.505 ± 0.052 & ... & L4 \\
        00302476+2244492 & 2 & 00:30:25 & +22:44:47 & 14.586 ± 0.036 & 13.975 ± 0.047 & ... & L0.5 \\
        00304384+3139321 & 2 & 00:30:44 & +31:39:31 & 15.480 ± 0.052 & 14.617 ± 0.057 & L2 & L3 \\
        00361617+1821104 & 2 & 00:36:16 & +18:21:10 & 12.466 ± 0.025 & 11.588 ± 0.028 & L3.5 & L4 \\
        $\vdots$  & $\vdots$ & $\vdots$ & $\vdots$  & $\vdots$       & $\vdots$ & $\vdots$ & $\vdots$\\
    \enddata
\end{deluxetable*}
\end{table}

\begin{figure}[h]
    \begin{center}
        \includegraphics[width=\columnwidth]{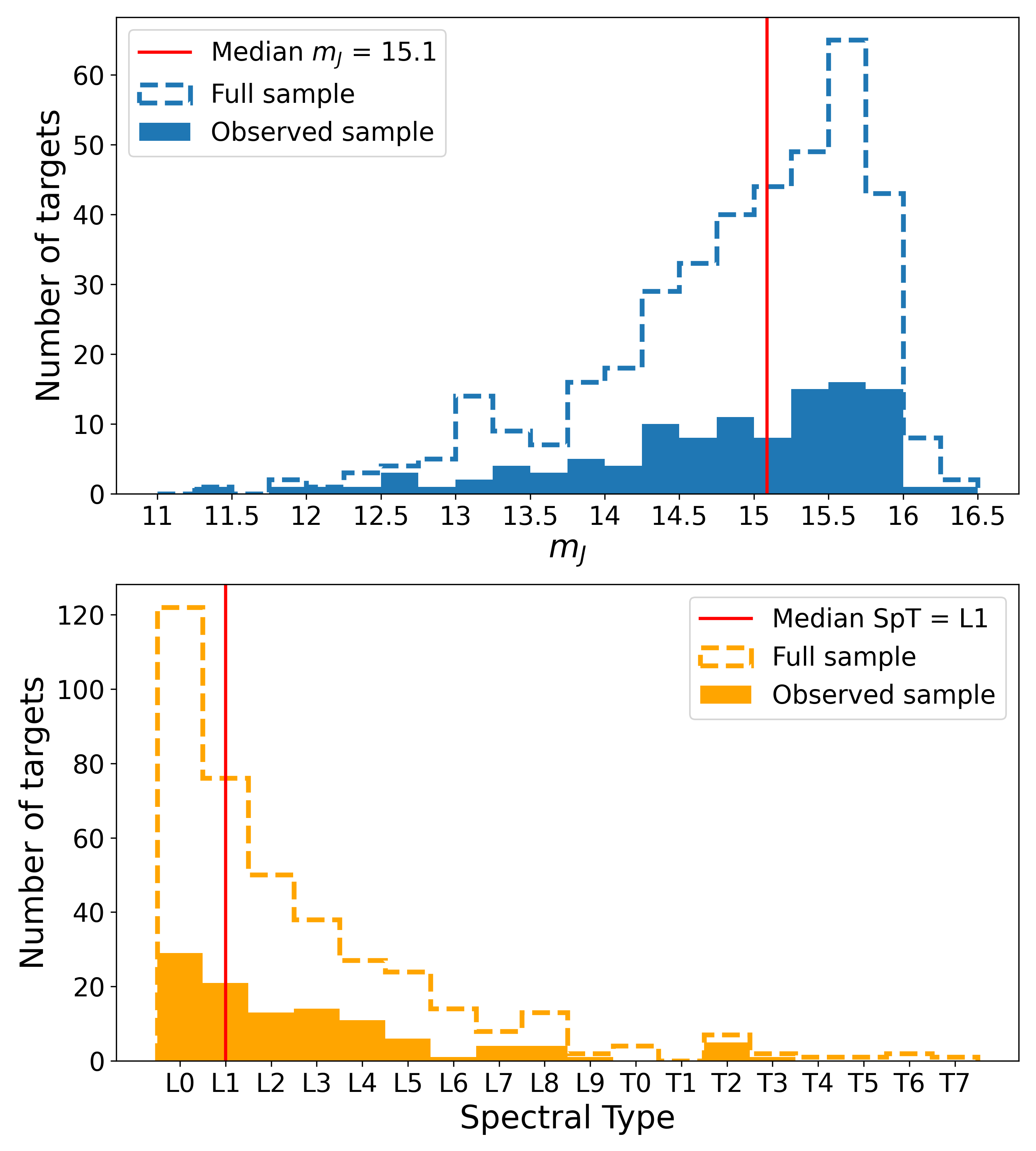}
        \caption{\textit{Top}: Histogram of apparent \textit{J}-band magnitudes of targets in the PINES sample. \textit{Bottom}: Histogram of spectral types.}
        \label{fig:histograms}
    \end{center}
\end{figure}

Histograms of the magnitudes and spectral types of targets are shown in Figure \ref{fig:histograms}. The full 393-target sample is shown as a dashed line, while the targets that have been observed as of October 2021 are filled in. Because the PINES sample is magnitude-limited, it is biased towards early-type L dwarfs, with 73\% of our targets being L3 or earlier. The median 2MASS \textit{J}-band apparent magnitude of the sample is $m_J = 15.1$, and the median spectral type is L1.

\begin{figure*}[ht]
    \begin{center}
        \includegraphics[width=\textwidth]{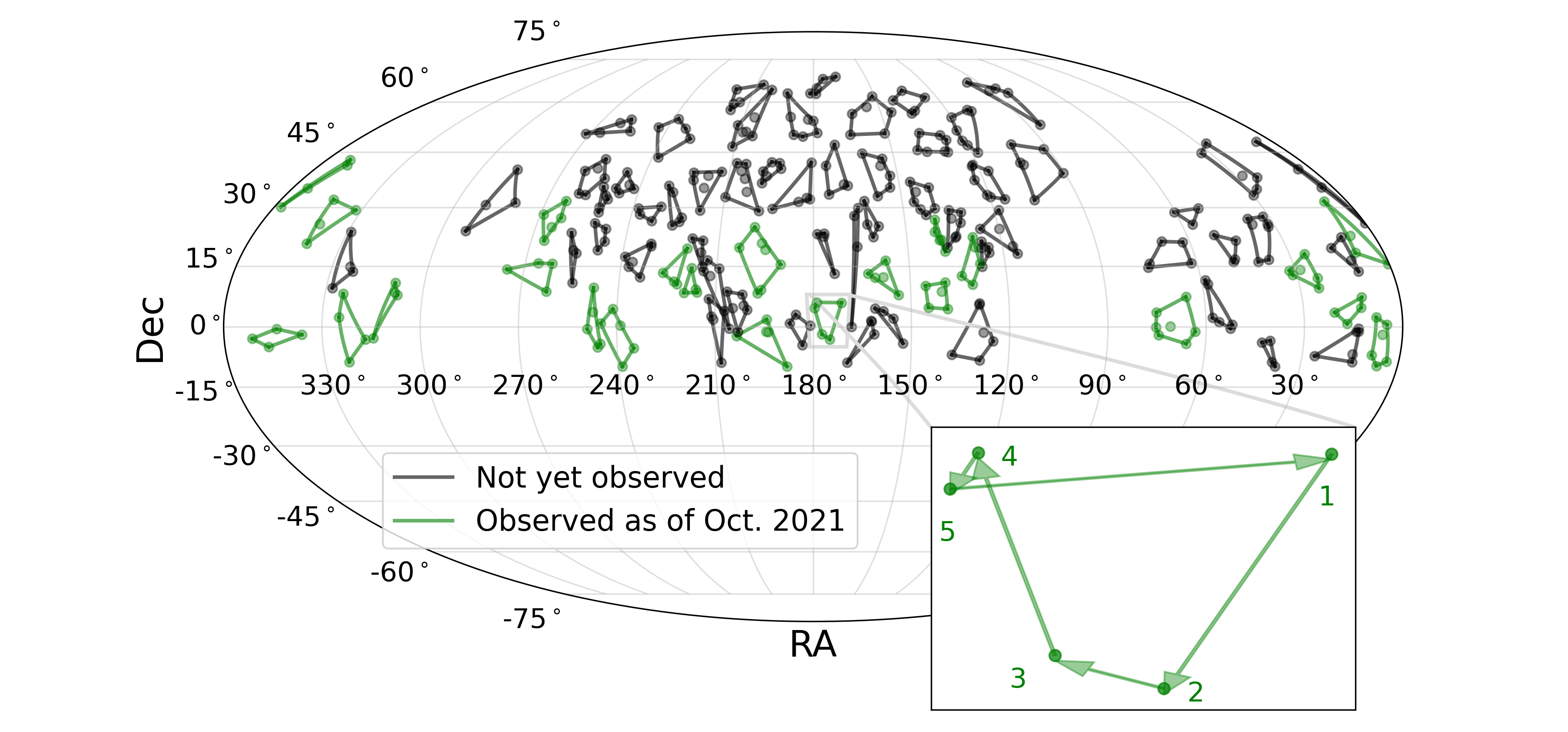}
        \caption{A sky map of the PINES sample. Targets that have been observed as of October 2021 are shown in green, while those that have yet to be observed are shown in black. Individual targets are shown as filled circles, while groups of targets are outlined with solid lines. The inset shows a zoom-in on one group and visualizes the order in which its members were targeted during PINES observations.}
        \label{fig:skymap}
    \end{center}
\end{figure*}

A sky map of the sample is shown in Figure \ref{fig:skymap}. PINES targets are distributed uniformly across the sky, except near R.A. 90$^\circ$ and 300$^\circ$, where the galactic plane inhibits the detection of faint L and T dwarfs. Targets in the sample were divided into groups using a clustering algorithm, which are represented by the solid lines in Figure \ref{fig:skymap}. We designed these groups such that the targets in each were separated by less than 15$^\circ$ on the sky, to reduce average slew times between group members.

\subsection{Observational Strategy}
\label{sec:observational_strategy}

PINES uses the observing strategy detailed in \citet{Tamburo2019}, in which a group of four to seven targets is repeatedly observed throughout an observing night (see the inset in Figure \ref{fig:skymap}). Each target is observed for an equal amount of time, and cycles are designed to take a maximum of one hour to capture short-duration transit events. This results in non-continuous time coverage for PINES targets, but transits can be detected by significant decrements of individual ``blocks'' of data. Groups of targets are scheduled for five nights of observations, which was found to maximize the number of observed transits in \citet{Tamburo2019}. This duration heavily biases the survey towards detecting short-period planets, and the same simulation showed that the vast majority of transits will belong to planets with orbital periods of 10 days or less. 

The strategy of cycling between multiple targets has been employed successfully in the search for transiting planets by other surveys \citep[e.g., MEarth,][]{Nutzman2008}, but as noted in \citet{Delrez2018}, a staring procedure (as employed by the SPECULOOS survey) maximizes signal-to-noise. This is indeed a limitation of cycling observations, as slews interrupt time that would otherwise be spent obtaining exposures. Additionally, cycling observations are at greater risk of systematics from flat fielding errors, as fields are generally not placed on the exact same pixels following slews. In the following section, we discuss our efforts to address the limitations of the cycling strategy.

\subsection{Guiding Performance}
\label{sec:guiding}
The PTO has a built-in off-axis auto-guider which can be used to keep sources on the same pixels over a set of exposures. Stable field positioning is important for accurate IR photometry, as image motions coupled with flat fielding errors and intra-pixel sensitivity variations can inject spurious signals into the final light curves \citep[e.g.,][]{Ingalls2012}. However, the existing auto-guider is inadequate for PINES observations because of the time overhead that it introduces. Following each slew, the observer must shift the field to the desired detector position, a task that can take several minutes to complete. Next, they must identify a suitable guide star and initiate the guiding system, which also requires about a minute of active user involvement. Because the PINES cycling strategy necessitates dozens of slews over the course of a night, the use of the existing PTO auto-guider would be prohibitively time-consuming and prone to human error. For this reason, we designed a custom guiding procedure for PINES that is quick, automatic, and places targets on roughly the same pixels throughout observations.

\begin{figure*}[ht]
    \begin{center}
        \includegraphics[width=\textwidth]{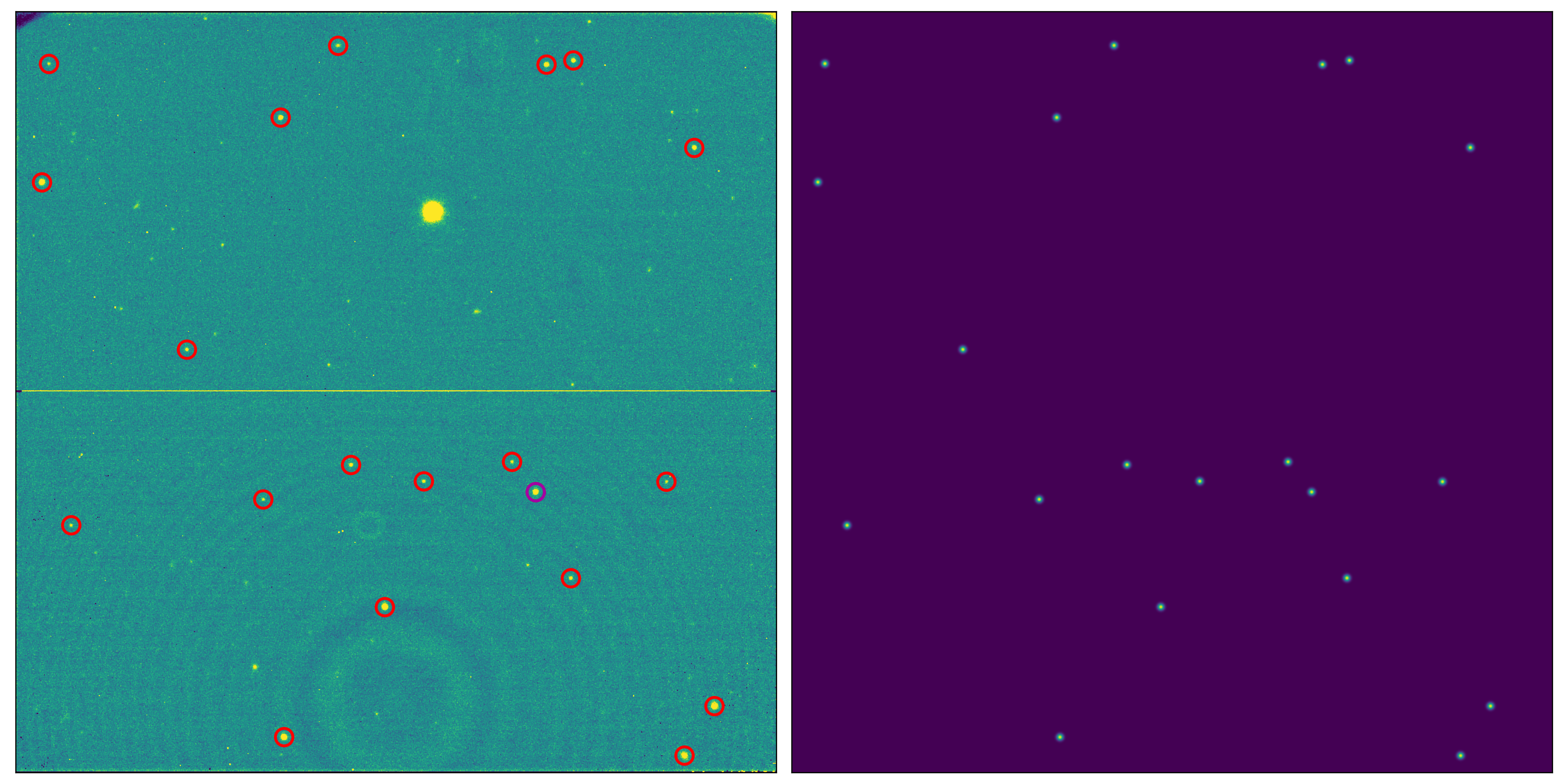}
        \caption{\textit{Left}: An example of a guide field image, with guide stars indicated with circles (with purple indicating the target and red indicating other sources). Bad pixels in this image have been interpolated with a 2D Gaussian kernel to improve source detection efficiency. \textit{Right}: The guide field synthetic image for this field.}
        \label{fig:guide_synthetic}
    \end{center}
\end{figure*}

At the beginning of each observing run, we position each field such that the target is placed on a region of the detector that is known to be free of bad pixels (see Figure \ref{fig:mimir_image}). We acquire a ``guide field" image at this location, and detect sources in this image with the \texttt{DAOStarFinder} routine from the \texttt{photutils} Python package \citep{photutils}. We use the detected source positions to create a synthetic guide field image, which consists of 2D Gaussians at the measured source positions (see Figure \ref{fig:guide_synthetic}). We use the synthetic guide field image in subsequent positioning calculations to measure the shifts that the telescope needs to execute to achieve the guide field position. A synthetic image is used to prevent image features like the crack seen in Figure \ref{fig:mimir_image} from dominating the fast Fourier transform (FFT) convolution with subsequent images.

\begin{figure}[hb]
        \includegraphics[width=\columnwidth]{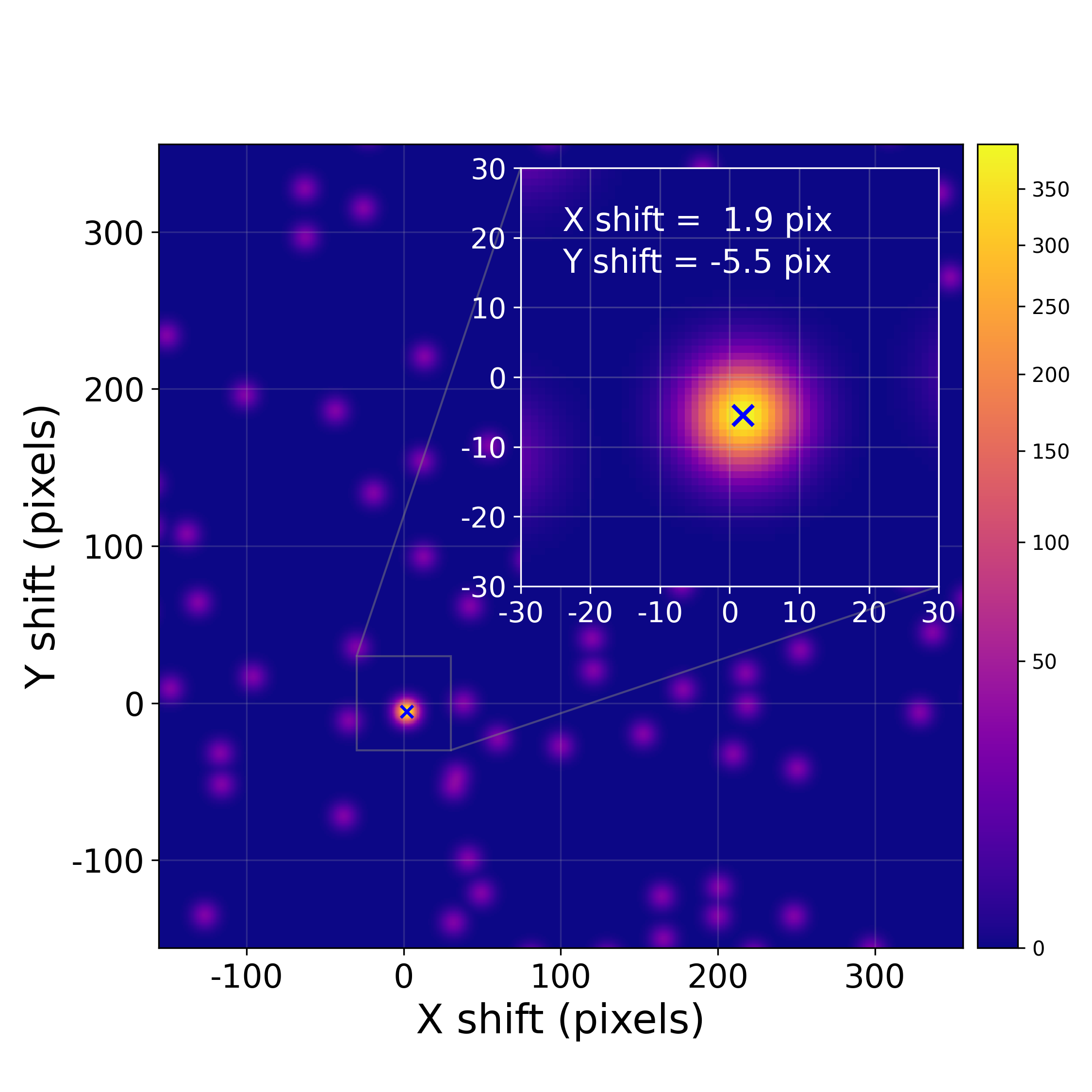}
        \caption{The FFT convolution of a science field synthetic image and a guide field synthetic image, which is used to determine the shifts needed to correct the PTO's pointing during PINES observations.}
        \label{fig:corr}
\end{figure}

When a new science image is saved, the guiding program first determines which field is being imaged by examining the telescope's reported right ascension and declination coordinates. It then performs a source detection on the newly written image and creates a synthetic science field image using the measured source positions. Subsequently, an FFT convolution is performed between the synthetic science field image and the synthetic guide field image, an example of which is shown in Figure \ref{fig:corr}. This figure shows many locations where the cross-correlation power is moderate, which is expected and is the result of image shifts that align a single pair of sources in the science image and the guide image. However, there is a clear peak in the cross-correlation power at +1.9 pixels in $x$ and -5.5 pixels in $y$. This is the shift that, when applied to this particular science image, would align it with the guide field position. The measured shifts are passed to the telescope to adjust its pointing back to the field's guide position. The entire process of measuring image shifts and adjusting the telescope position takes only a few seconds, so corrections are passed following every science image. 

\begin{figure}[t]
    \includegraphics[width=\columnwidth]{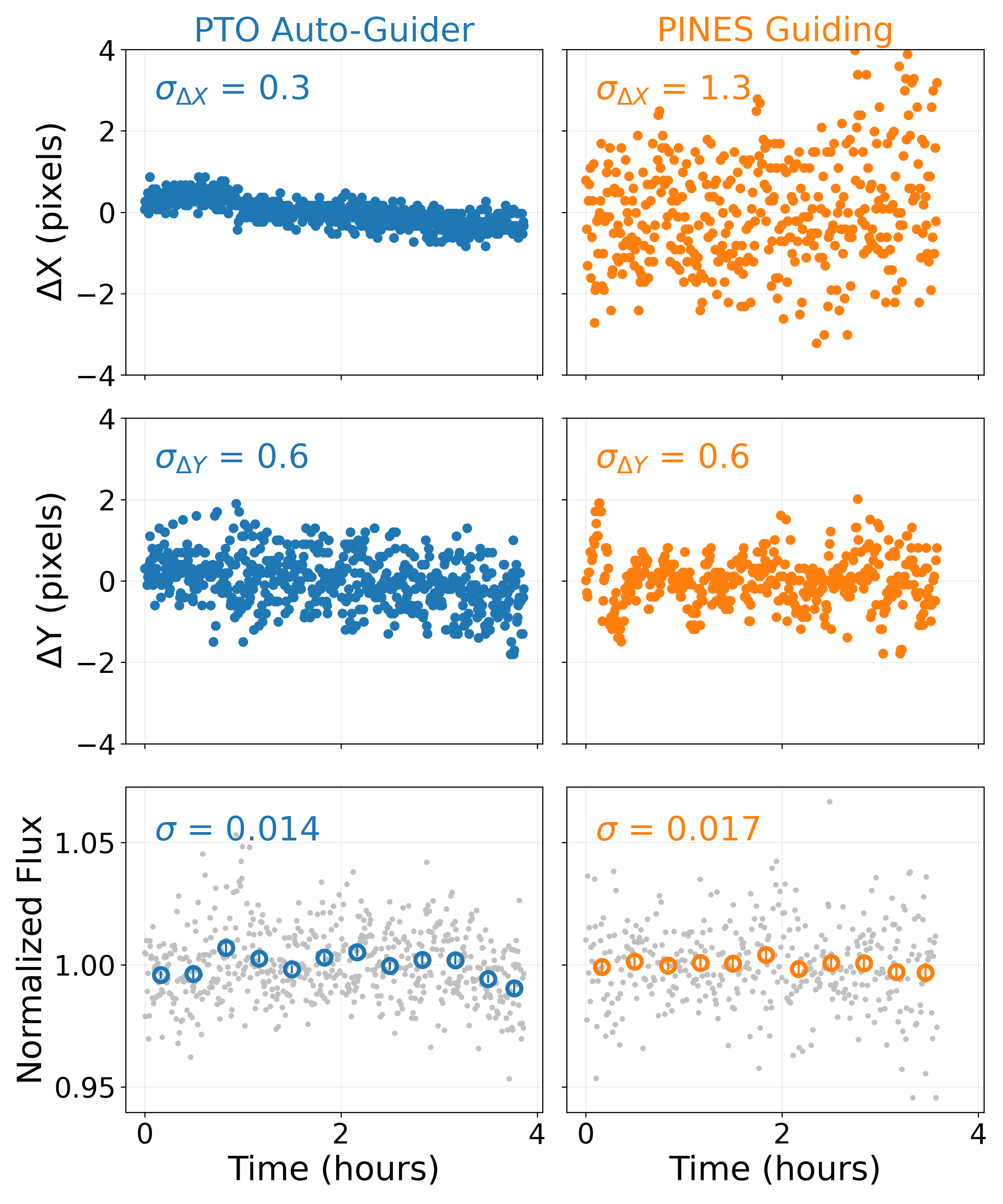}
    \caption{Comparison of the PTO auto-guider performance (left column) versus the PINES guiding system (right column). \textit{Top}: Measured $x$ pixel shifts. \textit{Middle}: Measured $y$ pixel (Dec axis) shifts. \textit{Bottom}: Light curves resulting from the two guiding approaches, which exhibit similar scatter.}
    \label{fig:pto_vs_pines_guiding}
\end{figure}

In Figure \ref{fig:pto_vs_pines_guiding}, we show a  comparison between the performance of the PTO auto-guider and the PINES guiding system. The top two rows show the $x$ and $y$ shifts (in pixels) between the measured field position and the guide position for about four hours on two different observing nights. On the first night, the target was observed with the PTO auto-guider, while on the second, it was positioned using the PINES guiding system. The PINES guiding observations have worse $x$ pointing precision, with a standard deviation of 1.3 pixels compared to 0.3 using the auto-guider. The $y$ pointing precision is comparable between the two guiders, however. We attribute the slightly worse pixel positioning from PINES guiding to the fact that pointing corrections are only issued on image readout (i.e., the correction frequency is set by the image exposure time, which is typically $\geq$ 30 s), whereas the PTO auto-guider passes corrections to the telescope about every second. Despite the worse pointing performance, light curves created using the two guiding approaches are essentially the same, as demonstrated by the standard deviations of the light curves in the bottom row of Figure \ref{fig:pto_vs_pines_guiding}. 

\subsection{Log Files}
\label{sec:log_files}
We automatically log the file name, date, target name, filter, exposure time, airmass, $x$/$y$ pixel shifts, and seeing FWHM of each image. The seeing estimate is given by the average FWHM of 2D circular Gaussian functions that are fit at the locations of detected sources in the image. These log files provide supporting information for the purposes of centroiding (Section \ref{subsec:centroiding}) and performing time-variable aperture photometry (Section \ref{subsec:aperture_photometry_and_bad_pixel_correction}). 

\section{The PINES Photometric Pipeline}
\label{sec:pipeline}
We process PINES data using a custom photometric pipeline to create light curves that can be searched for potential transit events. We call the pipeline the \texttt{PINES Analysis Toolkit} (\texttt{PAT}). We designed \texttt{PAT} to handle peculiarities of Mimir data that, if unaccounted for, would result in systematic errors in the final target light curves. It was also designed to function as automatically as possible, to facilitate the analysis of the large quantity of survey data. The software is available on GitHub\footnote{\texttt{PINES Analysis Toolkit} codebase: \url{https://github.com/patricktamburo/pines_analysis_toolkit}.} under an MIT License and version 1.0.0 is archived on Zenodo \citep{PINES_analysis_toolkit}. We detail the major steps of the pipeline in the following sections.

\subsection{Calibration Images}
\label{subsec:science_images_and_calibration_data}
Calibrations are typically taken once per run and consist of flat fields and dark exposures. Flat fields are obtained with a lamp-illuminated dome flat screen, with 100 lamp-on and 100 lamp-off images obtained in the same photometric filters as the science images. Combined lamp-on and lamp-off images are created by looping through each pixel, sigma clipping the 100 brightness measurements for that pixel with a 3-$\sigma$ clipping threshold, and taking the mean of the remaining measurements. We then construct a flat field by subtracting the combined lamp-off image from the combined lamp-on image, the values of which are then median-normalized to 1. 

Darks are constructed in a similar fashion, with sets of 10-20 individual dark exposures obtained using exposure times that match those of the science images. The same sigma clipping procedure used to create the flats is used to average the individual dark exposures together into a dark at each exposure time. A ``dark standard deviation" image is also produced for each exposure time, consisting of the standard deviation of the sigma-clipped values for each pixel. These capture the typical brightness fluctuations of detector pixels and can be used for bad pixel identification.

\subsection{Bad Pixel Identification and Data Reduction}
\label{subsec:bad_pixel_identification}
The flats and darks are used to identify variable, hot, and dead pixels on the detector. These pixels are spread more-or-less uniformly across the chip and produce unreliable flux measurements and would result in inaccuracies if included in a photometric aperture or background annulus. We therefore mask these pixels out and correct their values at a later stage in the pipeline.

First, a variable pixel mask is created for each science exposure time using the dark standard deviation images. Pixels with high values in these images vary significantly between one exposure and the next. Pixels that have standard deviations over five times the average standard deviation of the dark standard deviation image are flagged as variable. 

We then identify hot pixels using the dark images. Each pixel in a dark image is iterated over, and its value is compared to a box of its neighboring pixels. The pixel is flagged as hot if its value is significantly higher than the median of the neighboring pixels. Surrounding pixels can themselves be hot, in which case they will raise the standard deviation of the box and possibly prevent the identification of a hot pixel. The hot pixel identification is therefore performed as an iterative process, with previously identified hot pixels being excluded from subsequent calculations. Hot pixels are identified with a given box size and significance threshold until no more are found, at which point the box is shrunk and the process is repeated. Through testing, we found that the following combinations of box sizes and significance thresholds successfully flag almost all visible hot pixels: $13\times13$ pixels/10-$\sigma$, $11\times11$ pixels/10-$\sigma$, $9\times9$ pixels/10-$\sigma$, $7\times7$ pixels/9-$\sigma$, and $5\times5$ pixels/8-$\sigma$.

Dead pixel masks are created using the same iterative shrinking box approach, but flagging pixels if they have values significantly less than that of their neighbors in the flat images. Finally, we create a bad pixel mask (BPM) for each exposure time and filter combination using the variable, hot, and dead pixel masks. We dark subtract and flat field the data to reduce the raw science images, ignoring flagged pixels in the BPM.

\subsection{Target and Reference Identification}
\label{subsec:source_detection_and_astrometric_solution}
The reduced science images are used in all of the following steps of the pipeline, beginning with the identification of sources in the field that are used for photometry. In addition to the target L or T dwarf, we perform photometry on several reference stars (requiring at least three in every frame), which we use to remove flux changes that are common to all objects in the field. To choose a set of suitable reference stars, we first generate a list of sources in a user-chosen science image using the \texttt{DAOStarFinder} program from \texttt{photutils}. We identify the target from this list of sources based on its location on the detector, which is known \textit{a priori} from the positioning of the field during observations (see, e.g. Figure \ref{fig:mimir_image}). Suitable reference stars are then chosen, avoiding: 
\begin{enumerate}
    \item Those that have counts in the non-linear regime of the Mimir detector \citep[with pixel values $\gtrsim$ 4000 ADU,][]{Clemens2007}.
    \item Those that are much fainter than the target, and would introduce unnecessary noise to the target light curve. Typically, we set the dimness threshold to 0.3$\times$ the target's flux in the source detection image.
    \item Those that are near the edge of the detector, where shifts from guiding could move them off the chip.
    \item Those that are located close to other sources ($\lesssim 10$ pixels).
\end{enumerate}

On average, there are $24^{+10}_{-6}$ reference stars that meet these criteria in each field ($16^{th}-84^{th}$ percentile range). The large number of suitable reference stars generally allows us to perform differential photometry on our targets without inflating the scatter of the target's light curve, except for our brightest targets (see Section \ref{sec:photometric_performance}).

\subsection{Astrometric Solution and Source Spectral Types}
\label{subsec:astrometric_solution_and_source_spectral_types}

Mimir images do not have a built-in World Coordinate System (WCS) which can map $x$ and $y$ pixel coordinates to R.A. and Dec coordinates. We use a web API\footnote{\url{http://astrometry.net/doc/net/api.html}} to pass the source detection image to Astrometry.net \citep{Lang2010}, which solves for the WCS of the field. With the WCS data, we then convert the pixel locations of reference stars to R.A. and Dec coordinates. For every reference star in the field, we perform a 5"$\times$5" box search in Gaia Early Data Release 3 \citep{GaiaCollab2021} around its coordinates using the \texttt{Astroquery} Python package \citep{Ginsburg2019}, and save its ($G_{BP}-G_{RP}$) color. We also calculate the absolute \textit{G} magnitude of the reference using its measured parallax and discard references that are likely white dwarfs or giants. A sufficient number of dwarf stars are present in every field such that we can discard white dwarfs and giants as references without inflating the scatter of our final light curves, and we do so as a precaution against systematic effects that they could introduce through photometric variability or spectral mismatches with our targets. Finally, we use the ($G_{BP}-G_{RP}$) color to estimate the spectral type (SpT) and T$_{eff}$ of each reference star using a ($G_{BP}-G_{RP}$)-SpT relation, limited to M9V stars and earlier \citep{Pecaut2013}\footnote{Updated relations including the Gaia photometric bands are available online at \url{http://www.pas.rochester.edu/~emamajek/EEM_dwarf_UBVIJHK_colors_Teff.txt}}.

\subsection{Centroiding}
\label{subsec:centroiding}
Next, we measure the centroid locations of the target and reference stars in every image. For efficiency, we measure source centroids using small image cutouts (typically 16$\times$16 pixels), which we create using the positions in the source detection image and the pixel shifts from the observing logs. If there are known bad pixels in the source cutouts, their values are replaced with a 2D Gaussian convolution to improve centroiding performance (note that this replacement is not performed in the reduced image itself). We then apply a 2D Gaussian convolution to the entire cutout, which smooths out any remaining spurious pixel values (from, e.g., cosmic rays or unidentified bad pixels). We measure the centroid on the smoothed cutout with the \texttt{centroid\_2dg} function from \texttt{photutils}. The measured centroid in the cutout is then translated back into the full image pixel coordinates.

\subsection{Aperture Photometry and Bad Pixel Correction}
\label{subsec:aperture_photometry_and_bad_pixel_correction}
We perform simple aperture photometry with both fixed and time-variable apertures. We do fixed aperture photometry using circular apertures with radii of 3.0, 4.0, 5.0, 6.0, and 7.0 pixels, which encompass the range of typical FWHM seeing values at the PTO ($\sim$2"-3".5; the Mimir plate scale in wide-field imaging mode is 0".579/pixel). We also do time-variable aperture photometry, with aperture radii equal to 0.5$\times$, 0.75$\times$, 1.0$\times$, 1.25$\times$, and 1.5$\times$ a smoothed trend of the seeing FWHM.

\begin{figure*}[ht]
    \centering
    \includegraphics[width=\textwidth]{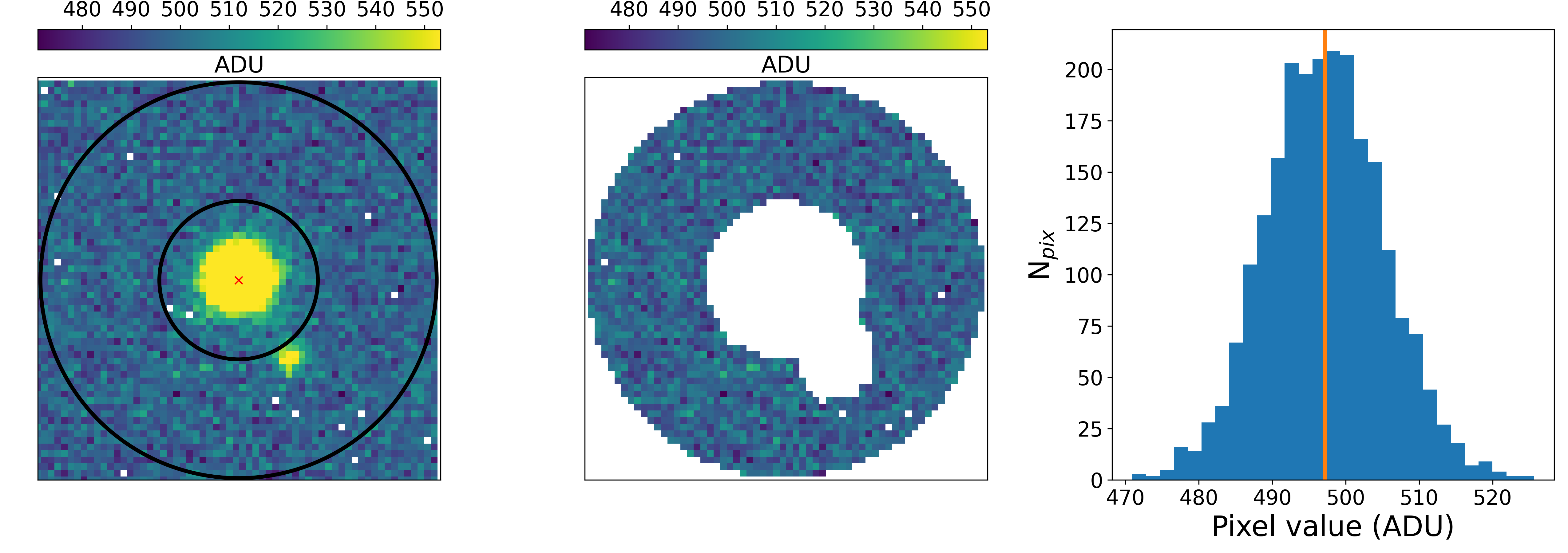}
    \caption{The background estimation process used for aperture photometry. \textit{Left}: A circular annulus is placed at the measured source position (red x), with bad pixels shown in white. \textit{Middle}: Pixels outside of the annulus (and if necessary, those containing sources) are ignored. \textit{Right}: A histogram of the resulting pixel values. The background is estimated as the sigma-clipped mean of these values (orange line).}
    \label{fig:background_estimation}
\end{figure*}

On average, we find that 2.7\% of the detector pixels are flagged as hot, dead, or variable in the BPM. Therefore, before we calculate the flux within an aperture, we check if it contains any known bad pixels. If it does, we fit a 2D Gaussian function to a cutout centered on the source position and replace bad pixels with the values from the fitted model. We purposefully position the target in a detector region that is mostly devoid of bad pixels, and as such it rarely requires a bad pixel correction. The reference stars, however, are frequently placed over bad pixels, and the values of these pixels have to be modeled to perform aperture photometry. The other possibility, of discarding reference stars that fall on bad pixels, is not feasible, as it would eliminate most (if not all) potential reference stars in any given field. 

We performed the following test to gauge the performance of our 2D Gaussian replacement procedure. We identified a source in one of our fields that has no bad pixels under its point spread function (PSF). We found the pixel that was nearest to the source centroid position on average, and calculated $\sigma_i$, its expected noise in frame $i$, using photon noise from the star and sky background, dark current, and read noise. We then created a new set of images, adding random noise to the value of the ``target" pixel by drawing from a normal distribution with $\sigma=\sigma_i$. We performed aperture photometry on this data set using a time-variable aperture with radius 1$\times$ a smoothed trend of the seeing FWHM. We then created another new set of images, this time flagging the target pixel as bad and replacing its value with our 2D Gaussian fitting procedure. We performed photometry on these images with the same aperture size.

Taking the ratio of the flux values measured from these two data sets, we found an average value of $1.0005\pm0.0029$. This suggests that our 2D Gaussian replacement procedure produces values that are consistent with the noise sources in our data. We experimented with a second-order correction to our 2D Gaussian approach, using the averaged residuals from Gaussians fit to all sources in an image to estimate updated replacement pixel values. However, we found no evidence of improvement over the first-order 2D Gaussian replacement, which is not surprising, as the previous test demonstrated that this approach is already consistent with the noise in the data.

With bad pixel values modeled with our 2D Gaussian procedure, the flux within the target aperture is summed up. We then estimate the contribution from sky background through a process that is illustrated in Figure \ref{fig:background_estimation}. First, a circular annulus centered on the source position is placed, with an inner radius of 12 pixels, and an outer radius of 30 pixels. Pixels outside of the annulus are ignored, as are pixels inside the annulus that contain any nearby sources (this is done with the \texttt{photutils} function \texttt{make$\_$source$\_$mask}). The remaining pixels are then sigma clipped with a threshold of 4-$\sigma$ to remove outliers. We then take the mean of the sigma-clipped values as the representation of the per-pixel sky background. This value, multiplied by the source aperture area, is subtracted from the source aperture sum to derive the background-corrected source flux. 

We tested for spatial inhomogeneities over scales comparable to the sky annulus by splitting 60$\times$60 pixel source cutouts (like the one shown in the left-hand panel of Figure \ref{fig:background_estimation}) into 15$\times$15 pixel sub-regions and compared the background value that we measure with the annulus approach described above versus the sigma-clipped means of the individual sub-regions (masking out the central source). In the several thousand cutouts that we tested, we found no evidence for a significant difference between the background value measured in the annulus versus the background measured in the sub-regions. We therefore conclude that spatial inhomogeneities in the background across the annulus and aperture should not be a concern for our background correction procedure.

\subsection{Light Curves}
\label{subsec:light_curves}
Finally, we create light curves using the background-subtracted fluxes from the target and reference stars. To remove trends that are common to all sources in the field due to changing observing conditions, an artificial light curve (ALC) is created using the weighted mean of reference star fluxes, in a procedure analogous to the one used by SPECULOOS \citep{Murray2020}. This process is intended to de-weight variable reference stars that could otherwise introduce spurious variability to the ALC-corrected target light curve. 

First, we restore the raw source fluxes from one of the aperture options described in Section \ref{subsec:aperture_photometry_and_bad_pixel_correction}. We normalize the raw target flux ($F_{\text{T}}$) using the weighted mean of the time series:

\begin{equation}
   \hat{F}_{\text{T}} = \frac
                                {F_{\text{T}}} 
                                { 
                                    \left[\frac{\sum_{j=1}^{N_{\text{frames}}} F_{\text{T}_j}W_{\text{T}_j}}
                                    {
                                        \sum_{j=1}^{N_{\text{frames}}}W_{\text{T}_j} 
                                    }\right]}
\end{equation}

Here, $W_{\text{T}_j} = 1/\sigma_{\text{T}_j}^2$, where $\sigma_{\text{T}_j}$ is the uncertainty on the target flux in frame \textit{j}, which is calculated using photon noise from the target, sky background, dark current, and read noise.

Next, the measured flux from the $i^{\text{th}}$ reference star ($F_{\text{R}_{i}}$) is read in, and normalized in the same way:

\begin{equation}
   \hat{F}_{\text{R}_i} = \frac
                                {F_{\text{R}_i}} 
                                { 
                                    \left[\frac{\sum_{j=1}^{N_{\text{frames}}} F_{\text{R}_{ij}}W_{\text{R}_{ij}}}
                                    {
                                        \sum_{j=1}^{N_{\text{frames}}}W_{\text{R}_{ij}} 
                                    }\right]}
\end{equation}

In this expression, $W_{\text{R}_{ij}} = 1/\sigma_{\text{R}_{ij}}^2$, where  $\sigma_{\text{R}_{ij}}$ is the uncertainty on the flux from the $i^{\text{th}}$ reference star in frame $j$. A weighted mean of the $\hat{F}_{\text{R}_i}$ is used to construct the ALC:

\begin{equation}
    ALC = \frac{\sum_{i=1}^{N_\text{refs}} \hat{F}_{\text{R}_i}W_{i}}{\sum_{i=1}^{N_\text{refs}} W_{i}}
    \label{eq:alc}
\end{equation}

$W_{i}$ is the total weight given to reference star $i$, the calculation of which is detailed in Appendix \ref{appendix:determining W_tot}. Assuming for the moment that the $W_i$ values are known and the ALC has been created, the corrected, differential target flux is then given simply by: 

\begin{equation}
    \hat{F}_\text{T}^\star = \frac{\hat{F_T}}{ALC}
\end{equation}

Next, we correct $\hat{F}_\text{T}^\star$ for linear correlations with airmass, seeing, and the target's centroid $x$ and $y$ pixel positions. We also include the target's intrapixel centroid location as a regressor, as near-infrared array detectors can exhibit sensitivity variations across single pixels that can impact differential photometry at the $\sim$1\% level \citep[e.g.,][]{Lauer1999, Ingalls2012}. Before the regression correction is performed, we calculate the Pearson correlation coefficient between the target flux and each regressor, along with the two-tailed p-value which tests for non-correlation. If the regressor's p-value is less than $0.01$, it is retained in the regression; otherwise, it is discarded. We then perform a linear regression fit between the significantly correlated regressors and the corrected differential target flux and divide the resulting fit from the target flux. Optionally, a linear or quadratic function can be fit to the corrected target flux to remove any remaining long-term trends throughout the night, but these options are turned off by default.

We create a target light curve for every photometry file described in Section \ref{subsec:aperture_photometry_and_bad_pixel_correction}. We then choose the best aperture by minimizing the average standard deviation of the corrected target flux evaluated over the duration of individual blocks of data. This is done for both nightly normalized light curves, where each night of data is normalized to 1, and globally normalized light curves, where the entire data set is normalized to 1. The globally normalized light curves allow us to search for variability trends with longer time baselines.

\subsection{Precipitable Water Vapor (PWV) Corrections}
\label{sec:pwv_effect}

A known limitation to the accuracy of differential photometry in the NIR is ``second-order extinction". Because the extinction coefficient depends on wavelength, stars with different SEDs will experience different amounts of wavelength-integrated extinction \citep{BailerJones2003}. By design, PINES targets are typically much redder than the reference stars in any given field, and as such are potentially susceptible to second-order extinction effects.

This does not pose an issue if the extinction is stable in time. Unfortunately, at NIR wavelengths extinction is primarily driven by PWV, which is notoriously time variable. Changing PWV levels can induce signals on the order of 1\% in target light curves at NIR wavelengths, and they can change rapidly enough throughout the night to mimic transit events \citep{Blake2008, Baker2017, Murray2020}. 

For this reason, we investigated the potential effect of second-order extinction on PINES observations, by examining the change in flux in the Mimir MKO \textit{J}- and \textit{H}-band filters as PWV levels are increased for different spectral types. We obtained NIR water vapor absorption spectra from ATRAN\footnote{\url{https://atran.arc.nasa.gov/cgi-bin/atran/atran.cgi}} \citep{Lord1992} ranging from 0 to 10 mm of PWV, and used low-resolution (R = 100) solar-metallicity BT-Settl models \citep{Allard2014} with $T_{eff}$ ranging from 1500 K to 6000 K to model the stars (1500 K is approximately the temperature of an L6 dwarf; 90\% of our targets have SpTs of L6 or earlier). For each combination of stellar spectrum, PWV spectrum, and filter bandpass, we interpolated onto a common wavelength grid, multiplied the three together, and integrated over wavelength to measure the total response. We then calculated $\Delta$F for each spectral type, the difference (in percent) between its normalized total response at a given PWV level and its response at a PWV level of 0 mm. 

\begin{figure}
    \centering
    \includegraphics[width=\columnwidth]{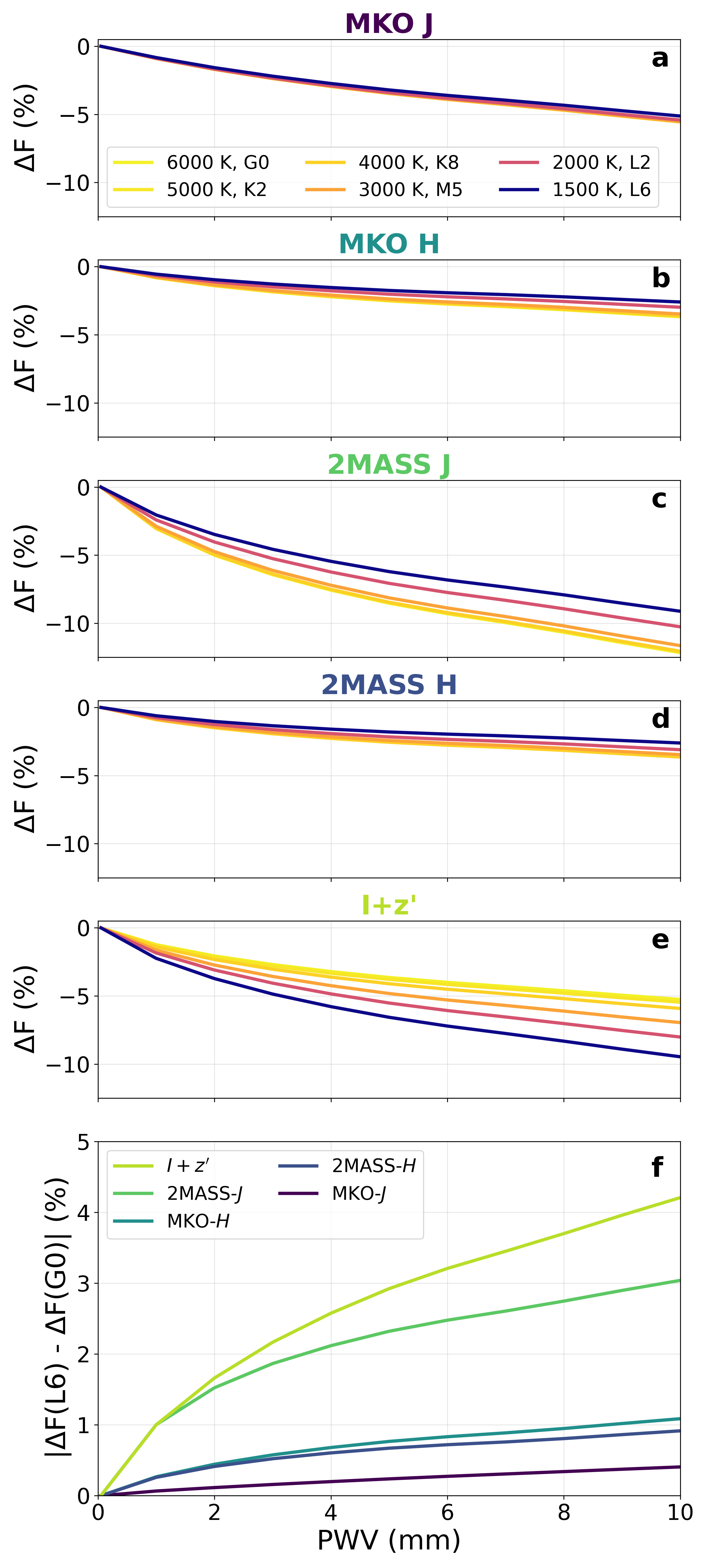}
    \caption{\textit{Panels a-e}: Theoretical response to changing PWV levels for different spectral types in various NIR filters. \textit{Panel f}: The magnitude of the PWV response of an L6 dwarf calibrated to the response of a G0 in each filter. The second-order extinction effect is minimized in MKO \textit{J}-band.}
    \label{fig:pwv_effect}
\end{figure}

The results of this analysis are shown in panels $a$-$e$ in Figure \ref{fig:pwv_effect}, along with calculations for three other NIR filters: 2MASS \textit{J}, 2MASS \textit{H}, and a model of the custom \textit{I+z'} filter used by the SPECULOOS transit survey \citep{Delrez2018}. The colored lines show the response of different spectral types to changing PWV in the respective band. Panels $a$-$e$ are shown with the same y-range, which reveals significant differences in the \textit{absolute} flux change of sources in different NIR bands. As PWV levels increase from 0 to 10 mm, sources in MKO \textit{J}-band will experience an average flux decrease of $6\%$, while those in MKO \textit{H}-band will decrease by just $3\%$. In 2MASS \textit{J}, 2MASS \textit{H}, and \textit{I+z'}, sources will undergo average flux decreases of $11\%$, $3\%$, and $7\%$, respectively. We note that our \textit{I+z'}-band results closely reproduce those from \citet{Murray2020}.

However, the relevant metric for second-order extinction is the \textit{relative} flux changes of sources in these various bands. This is illustrated in panel $f$ of Figure \ref{fig:pwv_effect}, which shows the magnitude of the difference in the response of an L6 target with that of a G0 reference star versus PWV. This panel demonstrates that second-order extinction is minimized in MKO \textit{J}-band, with a flux difference of just $0.4\%$ between an L6 and G0 over a change of 10 mm in PWV. In other bands, changing PWV can result in much larger flux mismatches: the effect is $0.9\%$ in 2MASS $H$-band, $1.1\%$ in MKO $H$-band, $3.0\%$ in 2MASS $J$-band, and $4.2\%$ in \textit{I+z'}-band. This result suggests that MKO \textit{J}-band is the best choice for minimizing systematic flux variations from second-order extinction effects, mainly because the filter was designed to avoid water vapor absorption lines \citep{Simons2002, Tokunaga2002}. For this reason, we have elected to perform PINES observations exclusively in MKO \textit{J}-band, contrary to the MKO \textit{H}-band recommendation in \citet{Tamburo2019}.

While the scatter of our \textit{J}-band light curves (see Section \ref{sec:photometric_performance}) is generally higher than the magnitude of potential second-order extinction effects, we implemented an optional procedure for correcting PINES light curves for changing PWV. Because the PTO does not have a direct way to measure PWV on-site, we obtain PWV estimates using the \texttt{Fyodor} Python package \citep{MeierValdes2021}. \texttt{Fyodor} utilizes publicly available vertical temperature and moisture profiles from Geostationary Operational Environmental Satellites (GOES) imaging data, which can be used to calculate PWV levels through the line-of-sight to a target. Typically, we find PWV changes of only a few mm throughout observing nights with \texttt{Fyodor}. With these values and the spectral types of our sources (see Section \ref{subsec:astrometric_solution_and_source_spectral_types}), we can estimate the expected response of stars in PINES images to changing PWV levels throughout observations and correct for it. However, because downloading GOES data is time-intensive and we expect the magnitude of second-order extinction to be small in MKO \textit{J}-band, this procedure is not enabled in \texttt{PAT} by default; instead, we use it as a check on light curves that show transit or variability signatures. Additionally, because the spectral types of our reference stars are known (see Section \ref{subsec:astrometric_solution_and_source_spectral_types}), the user can also limit the ALC creation to using only the reddest reference stars, instead of doing a full PWV correction.

\section{Photometric Performance and First Results}
\label{sec:performance}
We have applied \texttt{PAT} to observations of 83 PINES targets in \textit{J}-band, to date. In Section \ref{sec:photometric_performance}, we provide an assessment of the noise performance of the survey using the light curves of these targets. We also provide examples that validate the performance of \texttt{PAT} for the purpose of transit detection (Section \ref{sec:wasp2b}) and the identification of variable L and T dwarfs (Section \ref{sec:recovered_variables}). Finally, in Section \ref{sec:new_variables}, we identify a new variable L/T transition object: the T2 dwarf WISE J045746.08-020719.2.

\begin{figure}[b]
    \centering
    \includegraphics[width=\columnwidth]{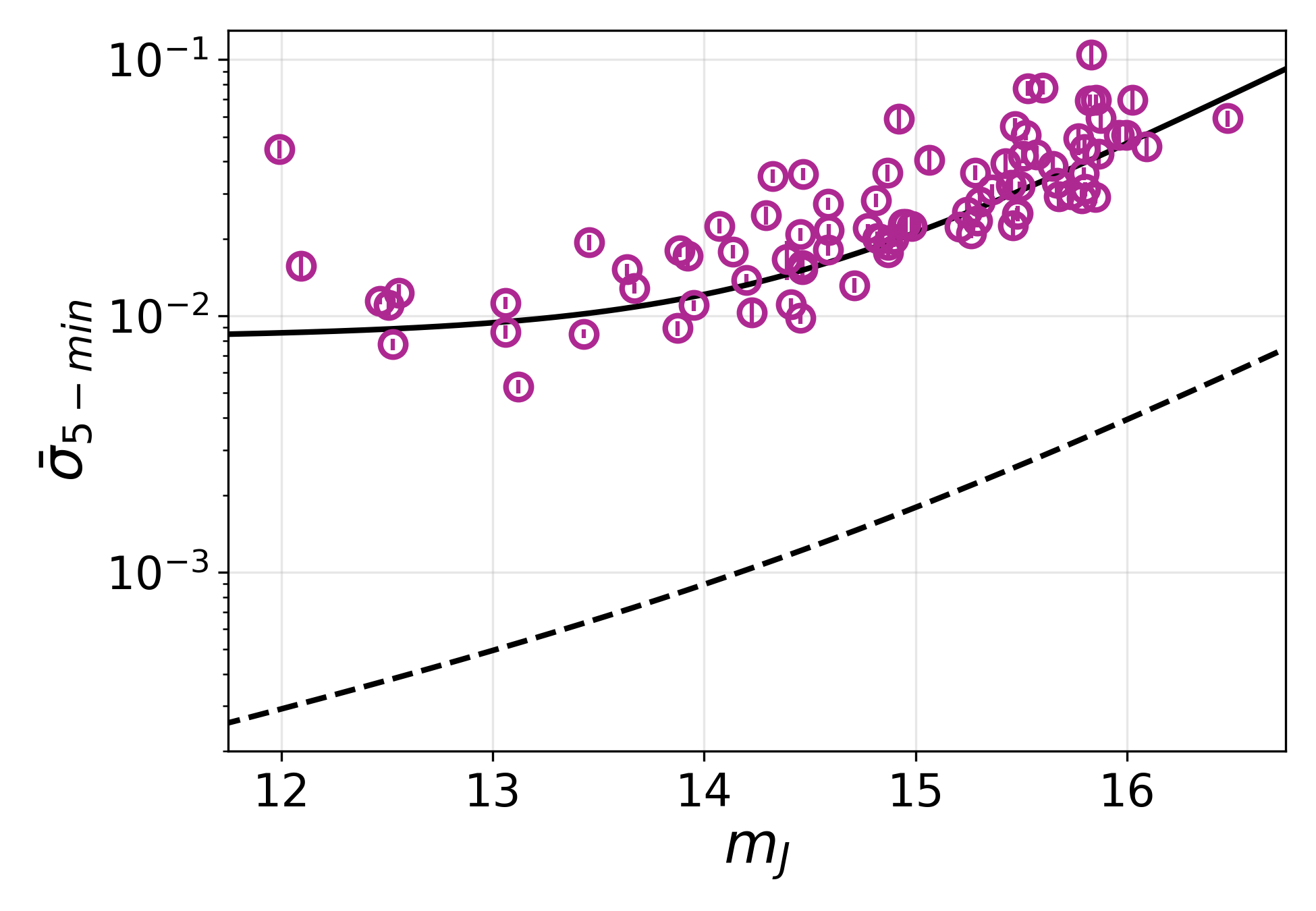}
    \caption{The average 5-min standard deviation ($\bar{\sigma}_{5-min}$) of PINES targets observed in \textit{J}-band (points with error bars). The dashed line indicates the expected noise performance from  \citet{Tamburo2019}, while the solid line shows an updated calculation using measured seeing, background, and throughput values, along with a logistic noise floor for targets brighter than $m_J \approx 14$.}
    \label{fig:noise_performance}
\end{figure}

\begin{table}[b]
    \centering
    \caption{Performance characteristics of the PINES survey, as of October 2021, compared to expected values from \citet{Tamburo2019}.}
    \begin{tabular}{l|r|r}
        \textbf{Parameter} &  \textbf{Measured} & \textbf{Expected}\\
        \hline
        Seeing & $2".6^{+0".5}_{-0".4}$ & $1".5$\\
        Nights requested & 207 & 220\\
        Nights assigned & 181 & 220\\
        Total night loss rate\footnote{This value accounts for whole-night losses from weather, forest closures, instrument failures, etc.} & $46\%$ & 30\%\\
        Weather loss rate\footnote{This value accounts for whole-night losses from weather only.} & 36\% & 30\%\\
        \textit{J}-band bkgd. (ADU/pix/s) & $17.9^{+8.0}_{-5.7}$ & 22\\
        \textit{J}-band throughput & 4.1\% & 35.0\%\\
    \end{tabular}

    \label{tab:year1}
\end{table}

\subsection{Photometric Performance}
\label{sec:photometric_performance}
\begin{figure*}[t]
    \centering
    \includegraphics[width=\textwidth]{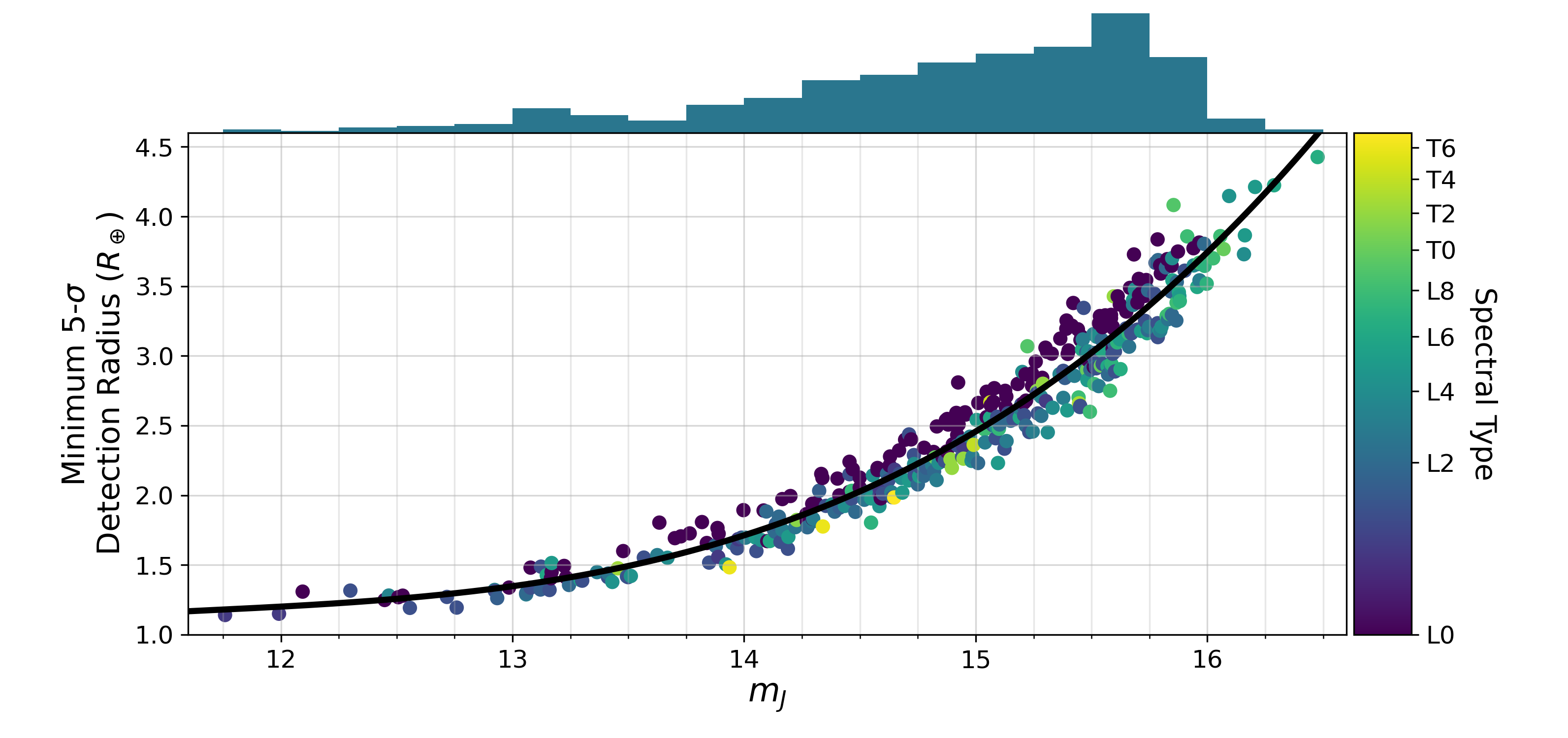}
    \caption{Estimated \textit{J}-band transit detection sensitivity for all of the targets in the PINES sample (colored points), assuming purely Gaussian noise as determined by the \textit{J}-band noise model shown in Figure \ref{fig:noise_performance}. Astrophysical or systematic noise sources, which are not captured in this model, will degrade this sensitivity estimation.} Targets in this panel are colored by their spectral type, and the black line shows a third-degree polynomial fit to the points. The histogram above the axis shows the \textit{J}-band magnitudes of targets in the PINES sample (the binning matches that in the top panel of Figure \ref{fig:histograms}). 
    \label{fig:transit_sensitivity}
\end{figure*}

In Figure \ref{fig:noise_performance}, we show the average standard deviation of the \textit{J}-band light curves analyzed with \texttt{PAT} over five minute intervals ($\bar{\sigma}_{5-min}$) . This time scale was chosen because all PINES objects are observed for at least five minutes in a single block, so the standard deviation over five-minute intervals can be evaluated for every block for every target. The expected noise performance from \citet{Tamburo2019} is indicated with a dashed line, including contributions from photon noise, sky background, dark current, read noise, and scintillation. The measured noise performance is worse than these expectations, and we attribute this mainly to poorer site seeing and lower net throughput than was anticipated. Our simulations assumed an average seeing FWHM value of 1".5, based on an optical site survey of the PTO location \citep{Tsay1990}. Instead, we have measured an average seeing FWHM of $2".6^{+0".5}_{-0".4}$, necessitating larger photometric apertures which result in higher contributions from sky background, read noise, and dark current. Additionally, we measure the net throughput in \textit{J}-band to be 4.1\%, compared to 35\% in \citet{Clemens2007}. The discrepancy could be due to degraded optical coatings in the instrument and/or degraded detector quantum efficiency.  We plan to address the throughput in a future instrument upgrade. If our measured throughput can be increased to the value measured by \citet{Clemens2007}, we could expect an improvement in our average light curve scatter by a factor of $\sim\sqrt{35/4.1}$.  Measured performance characteristics of the survey are given in Table \ref{tab:year1} in comparison to the expectations from \citet{Tamburo2019}.

The solid line in Figure \ref{fig:noise_performance} shows an updated calculation of the PINES noise performance using the measured seeing, background, read noise, and throughput values given in Table \ref{tab:year1}. Additionally, the noise model incorporates a logistic function to capture the noise floor that we find for targets brighter than $m_J \approx 14$. On average, there are fewer suitable reference stars available in the fields of bright targets. This tends to increase the noise of the ALC in these cases, which in turn inflates the scatter of the light curves of our brightest targets compared to expectations. 

We used the \textit{J}-band noise model in Figure \ref{fig:noise_performance} to estimate our transit detection sensitivity. We simulated cycling observations for every target in our sample assuming an exposure time of 60 seconds, a block length of 12 minutes, and a cycle length of 1 hour (typical of the PINES cycling strategy). We converted the known spectral types of our targets to $T_{eff}$ values using the $T_{eff}$-SpT relation from \citet{Faherty2016} for field M6-T9 dwarfs, then selected a random age for each from a uniform distribution from 0.25 to 10 Gyr. These temperatures and ages were used to obtain realistic radius estimates for each target (given the random age assumptions) using the evolutionary models from \citet{Baraffe2015}. We then calculated the minimum planet radius that could decrement one block of data to a significance of 5-$\sigma$, assuming the \textit{J}-band noise model shown in Figure \ref{fig:noise_performance}. We emphasize that this noise model does not include sources of astrophysical or systematic noise, which would strictly increase the size of the smallest detectable planet around a given PINES target. The results of this estimation, then, should be interpreted as lower limits to our planet detection sensitivity, and a full injection/recovery simulation will be required to establish our true sensitivity on a target-by-target basis.

We show the results of this analysis in Figure \ref{fig:transit_sensitivity}. The targets are colored by their spectral type, and a third-degree polynomial fit to the results is shown as a black line. A histogram of target \textit{J}-band magnitudes is shown on the top axis. A clear dependence on spectral type can be seen in this plot. Early L dwarfs are found in larger quantity above the polynomial fit than below, while the opposite is true for later-type sources. This is because sub-stellar objects shrink with time; if two objects have the same age, the earlier spectral type will have a larger radius than the later spectral type, requiring larger-radius planets to meet the 5-$\sigma$ detection criterion. 

\begin{figure*}[t]
    \centering
    \includegraphics[width=\textwidth]{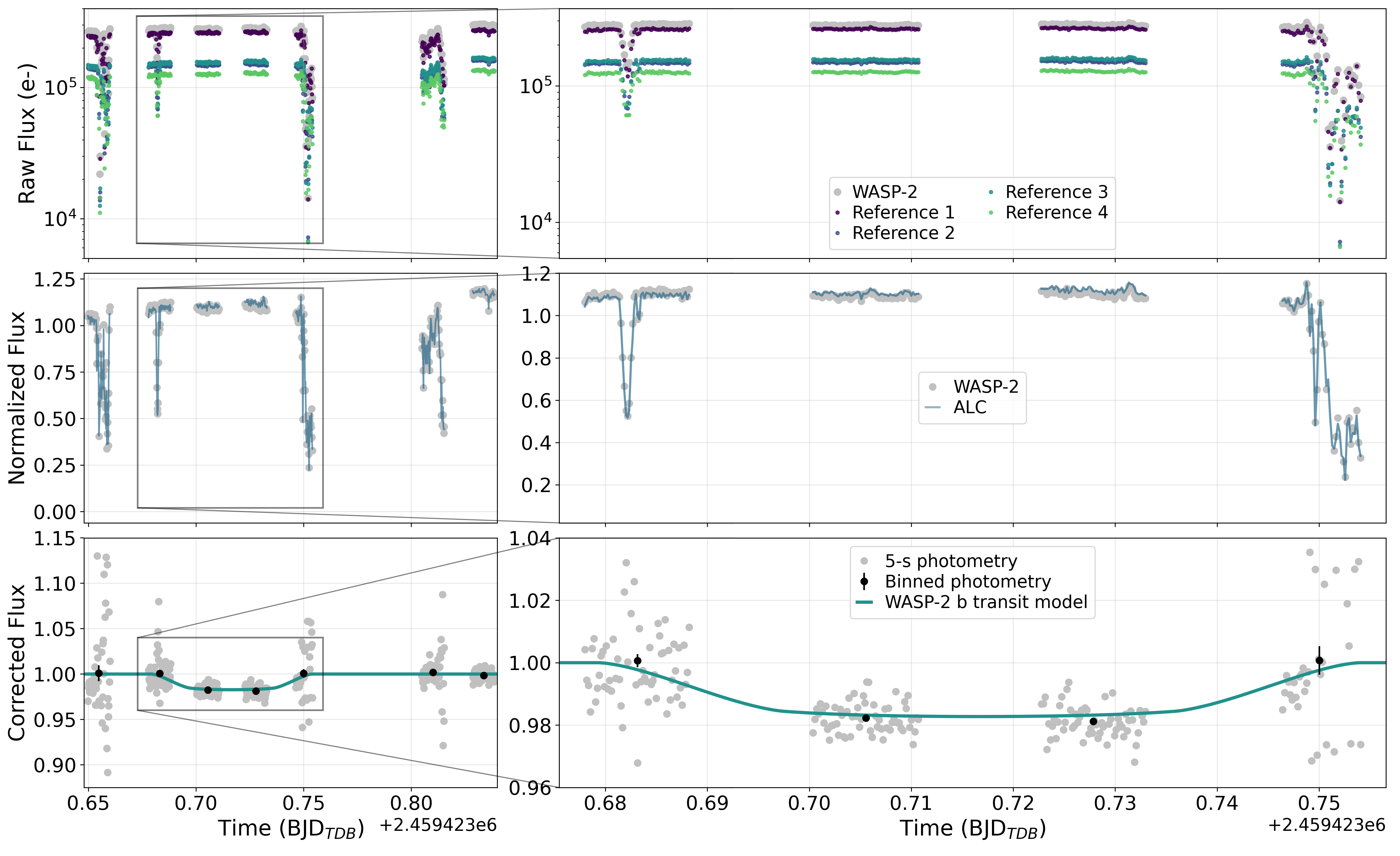}
    \caption{Transit recovery of WASP-2 b on UT 28 July 2021. The rows show different stages of \texttt{PAT}, going from raw flux to a final light curve. The left column shows the full night of data, while the right column shows a zoom in on the four blocks that cover the transit event. \textit{Top}: Raw flux measured for WASP-2 and four reference stars. \textit{Middle}: The normalized flux of WASP-2 and the ALC constructed from the weighted mean of the four reference stars. \textit{Bottom}: The normalized flux for WASP-2 corrected by the ALC and a linear baseline, along with a transit model of WASP-2 b.}
    \label{fig:wasp_2_b}
\end{figure*}

Under the assumption of purely Gaussian noise, this plot shows that  PINES is sensitive to the 5-$\sigma$ detection of planets smaller than Neptune ($R = 3.86$ $R_\oplus$) for all but the faintest targets in the sample. In particular, for the median \textit{J}-band magnitude of the PINES sample ($m_J = 15.1$), we find a lower radius sensitivity limit of $2.5\pm0.2$ $R_\oplus$. For our brightest targets, we calculate a lower sensitivity limit of 1.2 $R_\oplus$. We note that this lies within the 1-$\sigma$ uncertainty range of the ``Terran-Neptunian'' boundary at $R = 1.23^{+0.44}_{-0.22}$ $R_\oplus$ identified by \citet{Chen2017}, which marks the division between rocky and volatile-dominated planets. 

\subsection{Recovery of a Transit of WASP-2 b}
\label{sec:wasp2b}

In Figure \ref{fig:wasp_2_b}, we show an example of the light curve creation process for WASP-2, a binary star whose K1V primary is known to host a transiting 1.1 $R_{Jup}$ planet on a 2.15-day orbit \citep{CollierCameron2007, Daemgen2009}. We selected this target for PINES observations to test our ability to recover known transits with the analysis pipeline. WASP-2 is not an L or T dwarf, and its \textit{J}-band magnitude of 10.2 is about five magnitudes brighter than the median \textit{J}-band magnitude of the PINES sample. However, WASP-2 b's transit depth of 1.7\% is comparable to that of a super-Earth around a typical L or T dwarf (a 1.4 $R_\oplus$ planet would exhibit this transit depth around a 1.0 $R_{Jup}$ L or T dwarf), and for that reason, it serves as an appropriate test case for transit recovery.

We observed WASP-2 on the night of UT 28 July 2021 in \textit{J}-band with a 5-s exposure time. We switched between WASP-2 and one other target throughout the night, observing each in 15-minute blocks before switching to the other. Observing conditions were poor with intermittent clouds throughout the night, in some cases for extended periods which interrupted the desired cycle time of 30 minutes. Despite these interruptions, we managed to obtain seven blocks of data for WASP-2.

The top row of Figure \ref{fig:wasp_2_b} shows the raw flux measured for WASP-2 and four reference stars, using variable apertures with radii set to 1.5$\times$ a smoothed trend of the measured seeing. The short-timescale variability in the raw flux is due to time-variable cloud cover. The middle row shows the normalized target flux ($\hat{F}_T$) and the ALC, created using the weighted mean procedure described in Section \ref{subsec:light_curves}. The bottom row shows the target flux corrected by the ALC and a linear baseline, with a transit model of WASP-2 b overplotted. The corrected flux shows excellent agreement with the transit model, which validates the observing strategy and pipeline for the purpose of detecting $\sim$1\% transit depths.

\begin{figure*}[t]
    \centering
    \includegraphics[width=\textwidth]{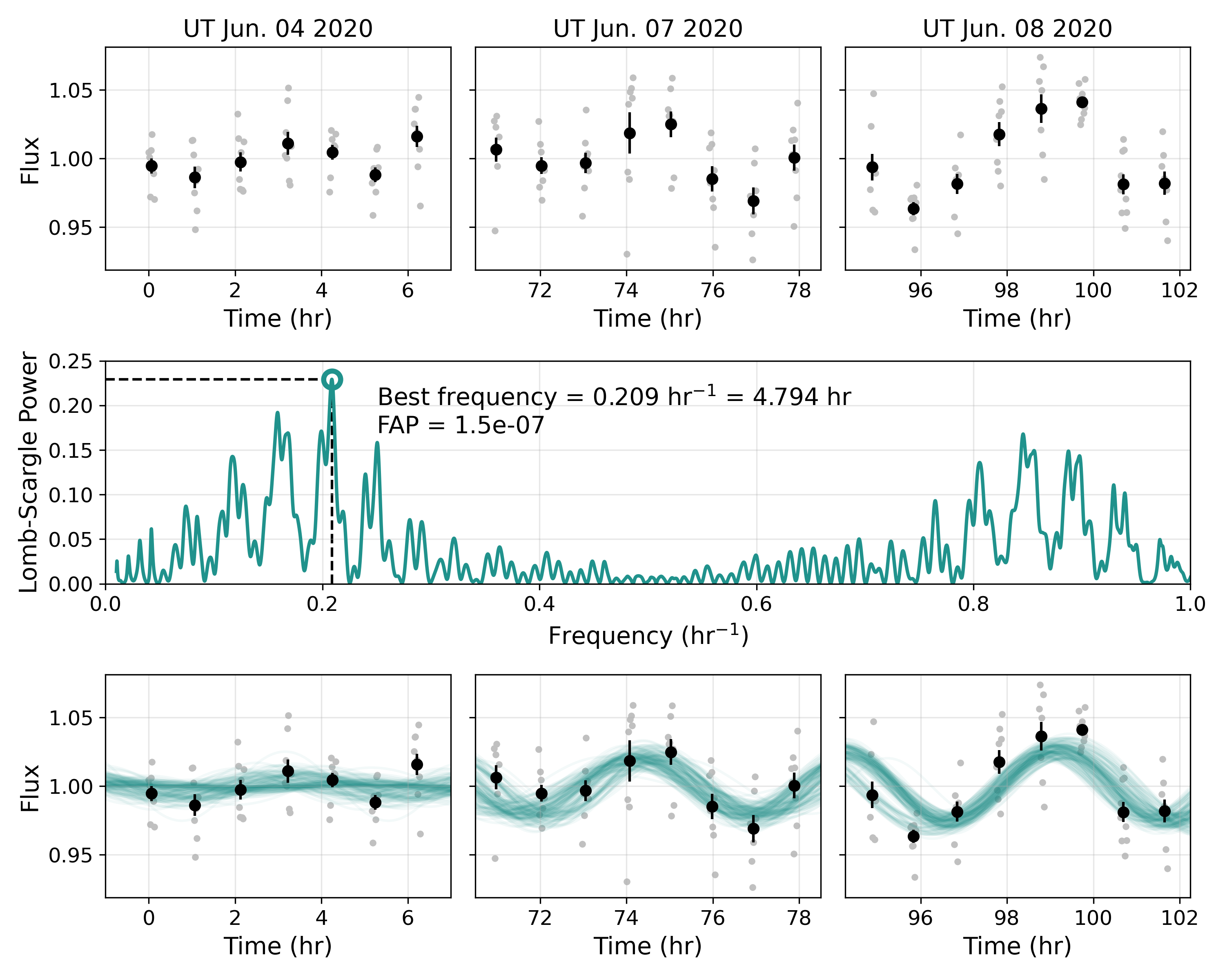}
    \caption{Variability recovery for SIMP 1629+0335. \textit{Top}: PINES observations taken in \textit{J}-band with a 60-s exposure time. The 60-s exposures are shown in grey, while the data binned over eight-minute blocks are shown in black with error bars. \textit{Middle}: A Lomb-Scargle periodogram of the unbinned data. \textit{Bottom}: Photometry with random MCMC models overplotted.}
    \label{fig:simp1629}
\end{figure*}

\subsection{Recovered Variables}
\label{sec:recovered_variables}
In this section, we describe the recovery of variability signatures from PINES targets that are known to be rapid rotators. Because of the discontinuous time coverage of PINES data, our observations are generally inadequate for the accurate characterization of the rotational variability of L and T dwarfs; rather, we show that we can readily \textit{detect} variability, and can use our observations to identify potential new variables which can be confirmed with follow-up observations (see Section \ref{sec:new_variables}). 

\subsubsection{SIMP J16291840+0335371}
\label{sec:simp_1629+0335}
SIMP J16291840+0335371 (SIMP 1629+0335) is a T2-type brown dwarf \citep{Deacon2011} which had a measured rotation period of 6.9 $\pm$ 2.4 hours and a peak-to-peak variability amplitude of 4.3 $\pm$ 2.4\% in \citet{Radigan2014}. SIMP 1629+0335 was observed by PINES for three nights in June 2020 in \textit{J}-band, and its light curve is shown in the top row of Figure \ref{fig:simp1629}. Images were taken with a 60-s exposure time, and the target was observed for eight minutes approximately every hour.

We created a Lomb-Scargle (LS) periodogram of the SIMP 1629+0335 light curve using the implementation in the \texttt{astropy} Python package with frequencies from 0.01 to 1 hr$^{-1}$ (equivalent to periods ranging from 1 to 100 hours). This periodogram, which is shown in the second row of Figure \ref{fig:simp1629}, shows a clear peak at $0.209$ hr$^{-1}$ ($4.8$ hr). We calculated the false alarm probability (FAP) of this peak using the approximation described in \citet{Baluev2008} and as implemented in \texttt{astropy}, and found it to be 1.5e-7. 

The period estimate from the LS periodogram is consistent with the rotation period measured in \citet{Radigan2014} to within 1-$\sigma$. To quantify the uncertainty on our measured period, we performed an MCMC simulation using \texttt{emcee} to fit our data with a sine curve of the form

\begin{equation}
    1+\frac{A_i}{2}\sin{(2\pi(\frac{t}{P}+\phi))} ,
    \label{eq:sine}
\end{equation}

\noindent where $A_i$ is the peak-to-peak amplitude on night $i$, $P$ is the period, and $\phi$ is the phase. We allow the fitted amplitude to vary from night to night because the variability of L/T transition objects can evolve significantly on short timescales \citep[e.g.,][]{Artigau2009, Gillon2013b}. The allowed periods were restricted to within $\pm$3 hours of the peak of the LS periodogram. We used 128 walkers for a total of 10000 steps, discarding the first 1000 steps of each chain for burn-in. Random samples from this MCMC simulation are shown in the third row of Figure \ref{fig:simp1629}. We find a period posterior that is bi-modal with peaks at $4.97^{+0.08}_{-0.14}$ and $6.22^{+0.13}_{-0.19}$ hr, both of which are consistent with \citet{Radigan2014}. The peak-to-peak amplitude posteriors reveal a slight increase in the variability from night-to-night: $1.24^{+1.08}_{-0.84}$\% on night 1, $4.18^{+1.02}_{-1.01}$\% on night 2, and $5.46^{+0.94}_{-0.89}$\% on night 3. 

The \textit{Spitzer} archive possesses time-series observations of SIMP 1629+0335, which we downloaded and analyzed to constrain the true period of this target. We analyzed data from two programs available in the archive -- PID: 80213 (PI: J Radigan) obtained in 2012 and PID: 11132 (PI: J Radigan) obtained in 2015. The 2012 dataset consists of $\sim24$~hr of continuous monitoring at $3.6~\mu$m, and the 2015 dataset consists of $\sim7$~hr of $4.5~\mu$m monitoring followed by $\sim7$~hr of $3.6~\mu$m monitoring. 

\begin{figure*}
    \centering
    \includegraphics[width=\textwidth]{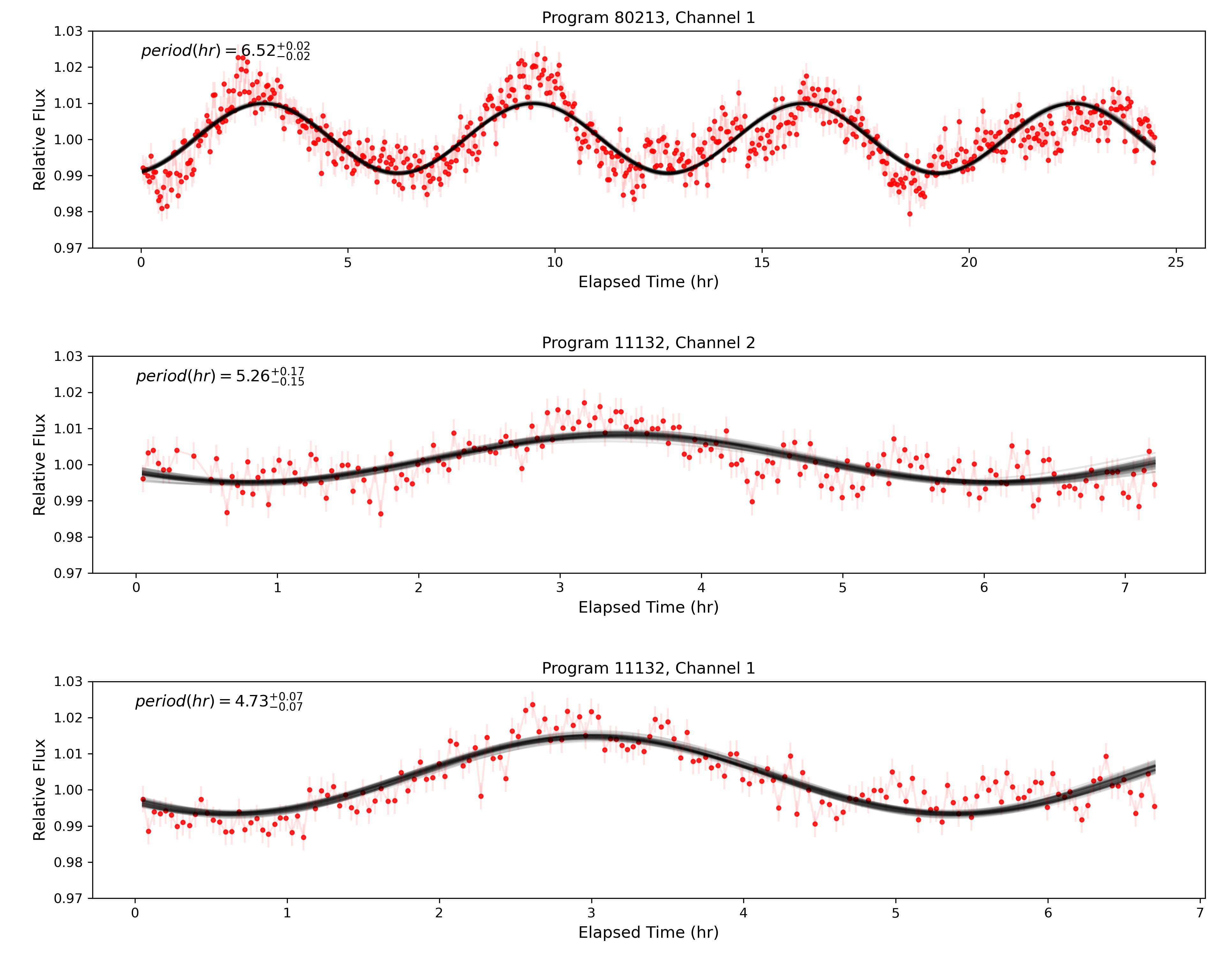}
    \caption{\textit{Spitzer} light curves for the variable brown dwarf SIMP 1629+0335.  The top panel displays the data taken in Channel 1 ($3.6~\mu$m) in 2012. The middle and bottom panels display the data from Channel 2 ($4.5~\mu$m) and Channel 1 ($3.6~\mu$m), respectively, taken in 2015. The corrected light curves are shown in red, with 50 random draws from our MCMC fit shown in black. The measured rotation period is significantly different in all three epochs.}
    \label{fig:SpitzerLCs}
\end{figure*}

We reduced and analyzed the three light curves following the methods presented in \citet{Vos2020} and briefly explain the steps here. We measured aperture photometry from the Basic Calibrated Data (BCD) images produced by the Spitzer Science Center. We used a variety of apertures ranging from $2.0-5.0$ pixels in steps of $0.25$, and ultimately choose the aperture resulting in the highest S/N final light curve. We corrected for the pixel-phase effect using a cubic function of the $x$ and $y$ coordinates, as described by previous light curve studies \citep[e.g.][]{Knutson2008, Metchev2015, Vos2020}, and find that this correction significantly reduces correlations between flux and pixel position for our target. We use two methods to check our light curves for significant variability -- a periodogram analysis method and a Bayesian method making use of the Bayesian Information Criterion (BIC). Both methods find that SIMP 1629+0335 appears variable at all three epochs. To measure the  rotation period of SIMP 1629+0335, we used \texttt{emcee} to fit a sinusoidal curve to each \textit{Spitzer} dataset. We use 1000 walkers for a total of 1000 steps and discard 100 burn-in samples from each chain. In Figure \ref{fig:SpitzerLCs} we show the three reduced light curves and their sinusoidal fits. We find that the measured rotation period changes significantly between the three different light curves: $6.52\pm0.02$ hr in Program 80213, $5.26^{+0.17}_{-0.15}$ hr in the first set of Program 11132 observations, and $4.73\pm0.07$ hr in the second set of Program 11132 observations. Programs 80213 and 11132 occurred three years apart, and a change in period over that time is not necessarily surprising. On the other hand, the two observations from program 11132 are separated by the time span of hours, representing two consecutive rotations of the brown dwarf. The change in the measured period is likely due to rapidly moving and/or evolving structures in the atmosphere of the dwarf, as have previously been noted in the light curves of variable, L/T transition brown dwarfs \citep{Apai2017}. The long-duration, continuous observations of SIMP 1629+0335 using \textit{Spitzer} suggest that the bi-modal period distribution measured from the PINES light curve is astrophysical in nature, and not a consequence of our observing strategy. 

Finally, it should be noted that inconsistent variability renders planet searching prohibitively difficult, as it is not possible to remove a signal that wholly accounts for the measured variability. For this reason, targets like SIMP 1629+0335 (and 2MASS 2139+0220, in the following section) are not searched for transiting planet signals in our pipeline.

\subsubsection{2MASS J21392676+0220226}
\label{sec:2mass_2139+0220}
\begin{figure*}
    \centering
    \includegraphics[width=\textwidth]{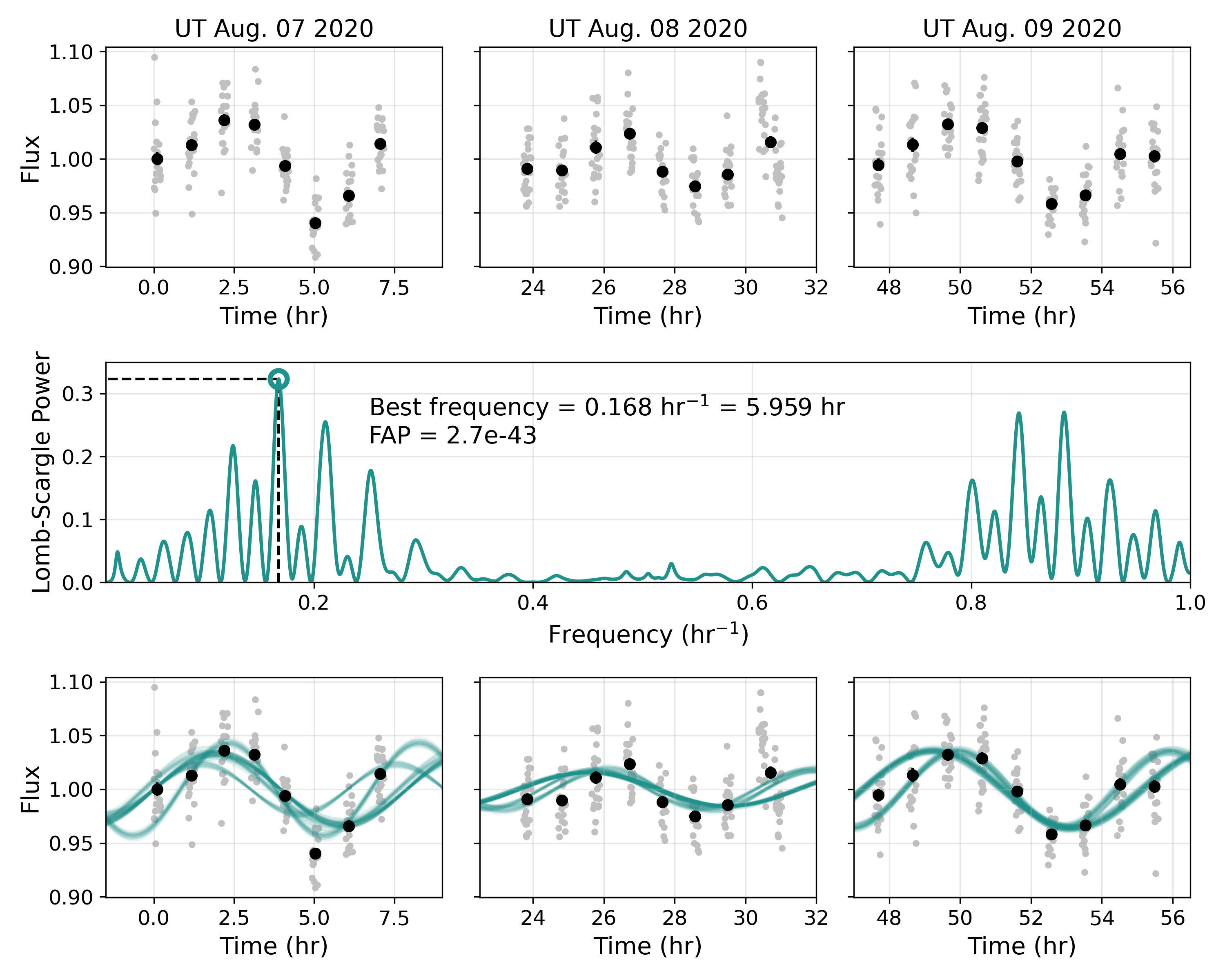}
    \caption{PINES light curve of the known variable 2MASS 2139$+$0220. The panels match those in Figure \ref{fig:simp1629}.}
    \label{fig:2m_2139}
\end{figure*}

2MASS J21392676+0220226 (2MASS 2139+0220) was found to be significantly variable in \citet{Radigan2012}, who measured a period of $7.721\pm0.005$ hr and peak-to-peak amplitudes up to 26\% in \textit{J}-band. We observed 2MASS 2139+0220 for three nights in \textit{J}-band in August 2020 and performed the same variability analysis as in the previous section. The results of this analysis are shown in Figure \ref{fig:2m_2139}.

In the LS periodogram for the 2MASS 2139+0220, the highest peak is at 6 hours, with an FAP of 6.7e-42. An MCMC fit with the function given in Equation \ref{eq:sine} reveals a multi-modal period distribution. The closest peak to the true period, at $7.91^{+0.05}_{-0.04}$ hr, is still significantly discrepant with the period measured in \citet{Radigan2012}, potentially because the variability of 2MASS 2139+0220 is known to not be perfectly sinusoidal (as assumed in our fitted model). Despite the period ambiguity, however, variability is clearly visible in the PINES light curve of 2MASS 2139+0220. Were this an unknown variable, our pipeline would have flagged it for follow-up observations, which could be performed at a more suitable cadence for characterizing the variability. 

\subsection{New Variable: WISE J045746.08-020719.2}
\label{sec:new_variables}
The previous two sections demonstrated our capacity for detecting variable objects in PINES data. Having performed these validation checks, we searched for new variables in our collection of PINES light curves and identified one previously unknown candidate variable: WISE J045746.08$-$020719.2 (WISE J0457$-$0207). \citet{Bihain2013} identified WISE J0457$-$0207 as a T2 dwarf using low-resolution NIR spectroscopy as part of a search for BDs in the solar neighborhood. \citet{Best2020} measured a parallax of $82.0 \pm 2.9$ mas for this object, with proper motion components $\mu_\alpha cos\delta = 93.0\pm3.0$ mas/yr, $\mu_\delta = -105.2\pm2.2$ mas/yr. We used these values to determine the probability that  WISE J0457$-$0207 is a member of nearby young associations with \texttt{BANYAN $\Sigma$} \citep{Gagne2018}, finding a 61.5\% probability for membership in the $\beta$ Pictoris moving group. We have no reason to suspect that this object is young \textit{a priori}, and an RV is required to determine WISE J0457$-$0207's potential $\beta$ Pictoris membership. A literature search for the target revealed that no light curve observations have been published for WISE J0457$-$0207.

\begin{figure*}
    \centering
    \includegraphics[width=\textwidth]{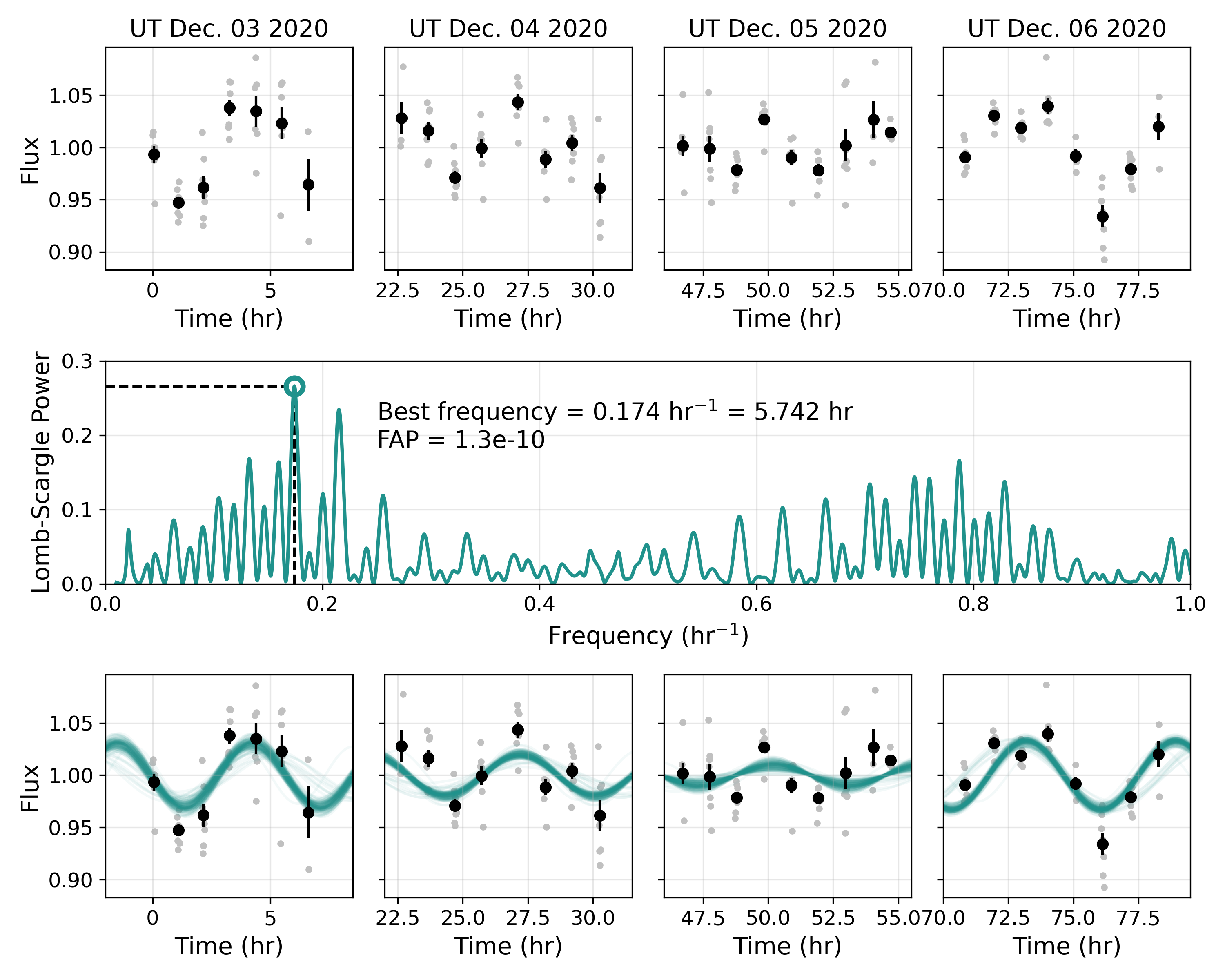}
    \caption{PINES light curve of the new variable WISE J0457$-$0207. We estimate its variability period to be 5.748 hours. The panels match those in Figure \ref{fig:simp1629}.}
    \label{fig:wise_0457}
\end{figure*}

We present the PINES light curve of WISE J0457$-$0207 in Figure \ref{fig:wise_0457}, which shows variability signatures consistent with those of the known variables recovered in Section \ref{sec:recovered_variables}. The target was observed for four nights in \textit{J}-band in December 2020 with a 60-s exposure time. We flagged the target as a potential variable after visual inspection, and then performed the variability retrieval analysis described in Section \ref{sec:recovered_variables}. The LS periodogram peak at 5.7 hr has a FAP of 1.3e-10. In contrast to the results for SIMP 1629+0335 and 2MASS 2139+0220, there is a single dominant peak in the period posterior for WISE J0457$-$0207, with a value of $5.748\pm0.011$ hours. The measured peak-to-peak variability amplitudes were $6.1\pm0.8\%$ on night 1, $4.0\pm0.6\%$ on night 2, $2.0\pm0.6\%$ on night 3, and $6.5\pm0.5\%$ on night 4, showing evolving variability that is typical of an L/T transition object \citep{Radigan2014, Metchev2015}. Some blocks of data are significantly discrepant with the pure sine model used in this analysis, however, and follow-up observations will be required to more fully characterize this object's variability.

\section{Conclusions}
\label{sec:discussion}
In this work, we provided an overview of PINES, a search for transiting satellites around a sample of almost 400 spectroscopically confirmed L and T dwarfs. Our main results can be summarized as follows:

\begin{enumerate}

    \item We described the observing facility, construction of the target sample, and observational strategy of PINES, which was designed to maximize the likelihood of detecting short-duration transits around L and T hosts.
    
    \item We detailed the custom guiding procedure that was created to execute the cycling strategy employed by PINES and compared the performance of this system to the built-in PTO auto-guider. We found that, despite worse pointing performance, light curves created with the PINES guiding systems have only marginally higher noise than those created with the auto-guider.
    
    \item We described the major steps of \texttt{PAT}, the custom photometric pipeline that we created to analyze PINES data. We showed how we identify bad pixels, reduce data, determine centroids, perform photometry, and create final target light curves using a weighted mean of reference star fluxes.
    
    \item We performed an analysis of the second-order extinction effect, in which changing PWV levels lead to different responses in the flux of different spectral types. We demonstrated that this effect is minimized in MKO \textit{J}-band, the filter used by PINES.
    
    \item We applied \texttt{PAT} to observations of 83 PINES targets. We used the resulting light curves to measure the noise performance of the survey and found that it was significantly higher than expectations from \citet{Tamburo2019}, due to a combination of worse average site seeing and net throughput. However, the derived \textit{J}-band noise model suggests that PINES will still be sensitive to the detection of sub-Neptune-sized planets around essentially the entirety of the sample.
    
    \item Finally, we detailed test cases that validate the performance of \texttt{PAT} for transit recovery and identifying variables. We identified one new variable in our observations thus far, the T2 dwarf WISE J045746.08$-$020719.2.
\end{enumerate}

The PINES survey is ongoing, with roughly 70\% of the sample remaining to be observed. PINES observations will permit stronger constraints on the occurrence rates of short-period planets around L and T dwarfs, which generally lie beyond the detection limits of previous optical exoplanet surveys. In turn, this effort will help to resolve the tension between competing lines of evidence for the existence of such planets around very-low-mass stars, brown dwarfs, and planetary-mass objects.

\begin{acknowledgements}
    The authors thank their anonymous referee for a thorough and insightful review that improved the quality of this work.

    This material is based upon work supported by the National Aeronautics and Space Administration under Grant No. 80NSSC20K0256 issued through the Science Mission Directorate.
    
    This work is based on observations made with the Spitzer Space Telescope, which is operated by the Jet Propulsion Laboratory, California Institute of Technology under a contract with NASA. The Spitzer data is available at the Spitzer Heritage Archive hosted by IRSA: \dataset[10.26131/IRSA430]{\doi{10.26131/IRSA430}}.
    
    Support for this work was provided by NASA through the NASA Hubble Fellowship grant HST-HF2-51447.001-A awarded by the Space Telescope Science Institute, which is operated by the Association of Universities for Research in Astronomy, Inc., for NASA, under contract NAS5-26555.
\end{acknowledgements}

\vspace{5mm}

\vspace{5mm}
\facilities{Boston University Perkins Telescope Observatory (Mimir), \textit{Spitzer Space Telescope} (IRAC)
}

\software{
    \texttt{Astrometry.net} \citep{Lang2010}, 
    \texttt{Astropy} \citep{astropy:2013, astropy:2018},
    \texttt{Astroquery} \citep{Ginsburg2019},
    \texttt{BANYAN $\Sigma$} \citep{Gagne2018},
    \texttt{BATMAN} \citep{Kreidberg2015}, 
    \texttt{emcee} \cite{emcee2013},
    \texttt{Fyodor} \citep{MeierValdes2021}
    \texttt{Matplotlib} \citep{Matplotlib2007},
    \texttt{NumPy} \citep{NumPy2020},
    \texttt{photutils} \citep{photutils}, 
    \texttt{PINES Analysis Toolkit} \citep{PINES_analysis_toolkit}
    }



\clearpage
\appendix
\section{Calculating Reference Star Weights}
\label{appendix:determining W_tot}

The reference star weights used to create the ALC in Equation \ref{eq:alc}, $W_{i}$, are calculated to give more weight to bright, stable reference stars while giving less weight to dim or variable reference stars. The weights are determined through an iterative process, based on a procedure described in \citet{Murray2020}. We construct a ``special" ALC (SALC) for every reference star in the field using the flux of all of the \textit{other} reference stars, weighted by their calculated uncertainties:

\begin{equation}
    \label{eq:salc}
    \text{SALC}_i =  \frac{\sum_{j=1, j \neq i}^{N_\text{refs}} \hat{F}_{\text{R}_j}W_{\text{R}_j}}{\sum_{j=1, j \neq i}^{N_\text{refs}} W_{\text{R}_j}}
\end{equation}

The SALC is then used to correct the flux of the reference star in question: 

\begin{equation}
    \hat{F}_{\text{R}_i}^\star = \frac{\hat{F}_{\text{R}_i}}{\text{SALC}_i}
\end{equation}

The standard deviation of $\hat{F}_{\text{R}_i}^\star$, $\sigma_{\text{R}_i}^\star$ is measured; then, $W_{i}$ is given by

\begin{equation}
    W_{i} =1/\sigma_{\text{R}_i}^{\star2}
\end{equation}

This operation is performed for every reference star, and then the process is repeated, with the $W_{\text{R}}$ values in Equation \ref{eq:salc} replaced by the calculated $W_{i}$ values in subsequent loops. This continues until the weights have stabilized to values within $1\mathrm{e}{-9}$ of their values in the previous loop. Once the weight values have converged, they are normalized to sum to one. 

We tested this weighting procedure using simulated photometry and found it to perform as expected. When we injected variability signatures into the reference stars, we found that those stars were down-weighted appropriately in the calculation of the target ALC, even if the variability was added to the brightest reference star.



\newpage

\bibliographystyle{aasjournal}
\bibliography{biblio}


\end{document}